\documentclass[aps,prd,amssymb,showpacs,floatfix,twocolumn,superscriptaddress,nofootinbib]{revtex4-1}
\usepackage{amsmath,amsfonts,graphics,epsfig,amssymb,bm}
\usepackage{multirow}
\usepackage{enumerate}
\usepackage{graphicx}
\begin{document}

\title{Testing CIBER cosmic infrared background measurements and axionlike particles with observations of TeV blazars}
\author{G. B. Long}
\email{longgb@mail2.sysu.edu.cn}
\affiliation{School of Physics and Astronomy, Sun Yat-sen University, Zhuhai, GuangDong, People's Republic of China}
\author{W. P. Lin}
\email{linweip5@mail.sysu.edu.cn}
\affiliation{School of Physics and Astronomy, Sun Yat-sen University, Zhuhai, GuangDong, People's Republic of China}
\author{P. H. T. Tam}
\email{tanbxuan@sysu.edu.cn}
\affiliation{School of Physics and Astronomy, Sun Yat-sen University, Zhuhai, GuangDong, People's Republic of China}
\author{W. S. Zhu}
\affiliation{School of Physics and Astronomy, Sun Yat-sen University, Zhuhai, GuangDong, People's Republic of China}

\begin{center}
\begin{abstract}
 The first measurements from the CIBER experiment of extragalactic background light (EBL) in near-infrared (NIR) band exhibit a higher intensity than those inferred through $\gamma$-ray observations. Recent theoretical-EBL intensities are typically consistent with the very
 high energy (VHE) $\gamma$-ray observations. Yet, it is possible that the excess NIR radiation is a new component of EBL and not in tension with the TeV spectra of distant blazars, since the hypothetical axion-like particle (ALP) may lead to a reduced opacity of the Universe for VHE $\gamma$-rays. In order to probe whether the excess component arises mainly from EBL,
thirteen observed spectra in high energy and VHE ranges from ten distant TeV BL Lac objects are fitted by four theoretical spectra which involve theoretical EBL (Gilmore ), Gilmore's EBL model including photon/ALP coupling, Gilmore's EBL with CIBER excess and the latter including photon/ALP coupling respectively.
We find the goodness of fit for the model with CIBER excess can be improved with a significance of $7.6~\sigma$ after including the photon/ALP coupling; Thus, the ALP/photon mixing mechanism can effectively alleviate the tension; However, the Gilmore EBL model, on the whole, is more compatible with the observed spectra compared to those with ALP, although individual blazars such as PKS 1424+240 and 1ES 1101-232 prefer the ALP-model. Our results suggest that the recent EBL models can solely
explain the VHE $\gamma$-ray observations, and assuming the existence of the ALP to alleviate the tension is not required in a statistical sense, thus the excess over the EBL models is less likely to be a new EBL component.
\end{abstract}
\end{center}
\maketitle

\section{Introduction}
\label{sec:intro}
The extragalactic background light (EBL) is the cosmic background photon field that is mainly composed by ultraviolet (UV), optical, and near-infrared (NIR) light. It is mainly produced by stars and interstellar medium in galaxies throughout the cosmic history. Consequently, the EBL is an important observable for models of galaxy formation and evolution \cite{Hauser2001}. The EBL could be directly measured with different instruments (e.g.\ Refs.~\cite{Sano2015, Sano2016,
Matsumoto2015, Tsumura2013c, CIBER2017}), but it is a challenge to accurately subtract the foreground of zodiacal light and diffuse Galactic light. The lower limits of the EBL can be estimated using deep-galaxy-surveys data (e.g.\ Refs.~\cite{Keenan2010}). Several empirical EBL models based on different complementary methodologies have been developed (e.g.\
Refs.~\cite{Franceschini2008, Dominguez2011, Kneiske2010,
Finke2010, Gilmore2012, Stecker2016}). Another technique to constrain the EBL indirectly is based on the $\gamma$-ray observations of extragalactic sources.

The very-high-energy (VHE, above 100~GeV) $\gamma$-rays
from distant ($z\geq0.1$) blazars would
suffer significant attenuation by
pair-production interactions with the EBL during the propagation in extragalactic space
\cite{Nikishov1962, Hauser2001, HESS2006, Dwek2013,
Costamante2013}. As a result, a $\gamma-\gamma$ absorption imprint of the EBL is carried on the VHE observed spectra of
blazars. This provide an independent way to constrain
indirectly the EBL intensities (for reviews see e.g.\
Refs.~\cite{Hauser2001, Dwek2013, Costamante2013}) and test
the empirical EBL models, e.g.\
Refs.~\cite{Franceschini2008, Dominguez2011, Kneiske2010,
Finke2010, Gilmore2012, Stecker2016}, by comparing the
difference between the assumed-intrinsic spectra and the
observed one. Thanks to the discovery of more and more distant Tev blazars, several groups have detected this imprint, e.g.\ Refs.~\cite{HESS2006, MAGIC2008, Fermi2012}, and found the intensities were near the galaxy-count lower limits.

 The spectra of a few individual blazars such as PKS1424+240 appear to be unexpectedly hard \cite{Abdalla2018}. Furthermore, hints for a reduced gamma-ray opacity in the form of a pair-production anomaly \cite{Horn2012} (i.e.,~VHE $\gamma$-ray observations require an EBL level below the lower limits from galaxy counts), or unusual redshift-dependent intrinsic (after correction for EBL absorption) spectral hardening found in large samples, have been claimed by several authors \cite{Essey2012, Rubtsov2014, Galanti2015, Korochkin2018}. However, no significant and systematic anomalies on the entire sample were revealed by recent systematic studies of spectra fitted with recent theoretical EBL absorption on large samples of VHE blazars, see e.g.,\ Refs.~\cite{Biteau2015,
 Dominguez2015, MAGIC2017, HESS2017, Zhong2018, Fermi2018, Abdalla2018, Desai2019, MAGIC2019}. The intensities of these recent EBL models (e.g.~Franceschini\ 2008 \cite{Franceschini2008}, Dominguez\ 2011 \cite{Dominguez2011}, Kneiske\ 2010 \cite{Kneiske2010},
 Finke\ 2010 \cite{Finke2010}, Gilmore\ 2012
 \cite{Gilmore2012}, Stecker\ 2016 \cite{Stecker2016}) are close to that from the galaxy counts. At present, many EBL measurements inferred by $\gamma$-ray observations reveal that its intensities are generally consistent with those given by the recent EBL models, see e.g.,~Refs.~\cite{Fermi2012,
 Gong2013, HESS2013, VERITAS2015, Biteau2015, MAGIC2016,
 Armstrong2017, Desai2019, Pueschel2019}.

\begin{figure}
\centerline{\includegraphics[width=1\columnwidth]{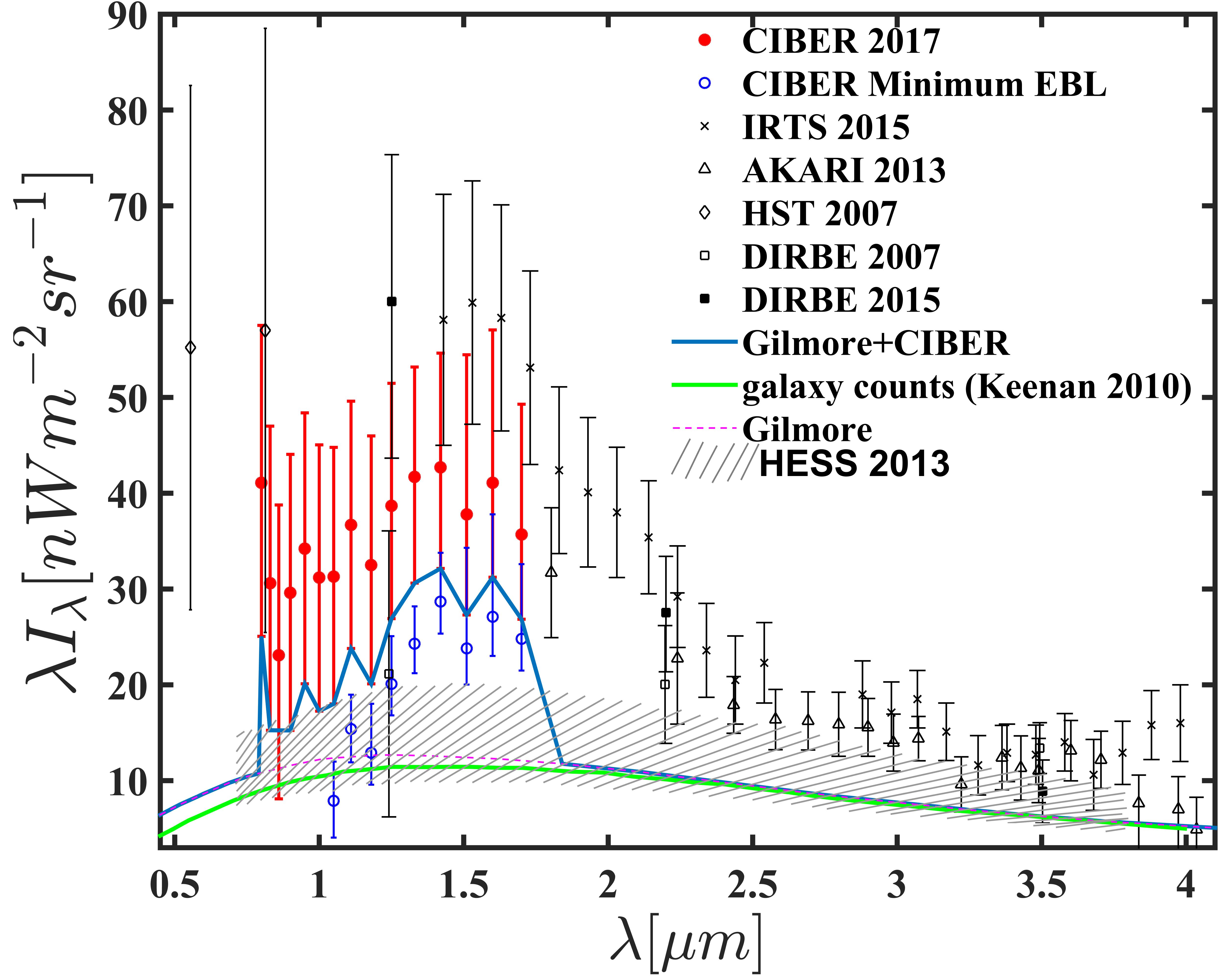}}
\caption{
 The EBL spectrum at NIR band. The data points represent the
 local measurements: nominal and minimum CIBER EBL (red filled circles and blue circles \cite{CIBER2017}), IRTS (crosses \cite{Matsumoto2015}),
 AKARI (open triangles \cite{Tsumura2013c}), HST (open
 rhombuses \cite{Bernstein2007}), COBE (filled and open
 squares \cite{Levenson2007, Sano2015}). Gilmore (carmine)
 means theoretically inferred spectrum in
 Ref.~\cite{Gilmore2012} and the data is obtained from
 online resources \cite{Gilmore2011}. The green curve is the
 EBL derived from deep galaxy counts
\cite{Keenan2010}. Gilmore+CIBER (blue curve) combine the
Gilmore EBL spectrum and the CIBER measurement value where
we take the lower limit of a $\sigma$ statistical
uncertainty and a systematic error (but we take a value of about 15.2\,$\rm nWm^{-2}sr^{-1}$ for the third data point since its lower limit below the value derived from deep galaxy counts). The hatched region gives
the $\gamma$-rays constraints by HESS~\cite{HESS2013}.
}
\label{fig:1}
\end{figure}

However, most of the direct local measurements of EBL from several
different groups (e.g.\ Refs.~\cite{Sano2015, Sano2016,
Matsumoto2015, Tsumura2013c, CIBER2017}) in NIR band
(0.8-5\,$\mu$m), especially the first result recently
released at 0.8-1.7\,$\mu$m by CIBER \cite{CIBER2017},
exhibit larger intensity than those predicted by EBL models
and measured using $\gamma$-rays observations, see Fig.~\ref{fig:1}. Moreover, the attenuation by pair producing interaction for the TeV
photons from distant blazars is most sensitive to the NIR
EBL intensity at the wavelength around 1\,$\mu$m
\cite{Kori2017}. Thus, even if
only taking into account the measurements of CIBER,
the excess radiation may lead to pair-production anomaly or an un-physically hard intrinsic spectrum (the spectrum is harder than that expected from the traditional acceleration and radiation model of blazar) for distant TeV blazars. If the excess radiation is a new EBL component, unconventional solutions are required to alleviate the tension between the VHE spectra of blazar and the measurements of CIBER.

To alleviate this tension, a possible attractive way is to  assume the existence of a very light pseudo-scalar
spin-zero boson called axion-like particle (ALP)
\cite{Kori2017, Kalashev2019}, which is predicted by many extensions of
the Standard Model of particle and especially superstring
theories (for reviews see e.g.\ Refs.~\cite{Masso2010}). ALP
is characterized by a two-photon coupling (a two-photon
vertex) \cite{Raffelt1988}, so it can mix with photons
in external magnetic fields. This mechanism could prevent a considerable fraction of VHE photons from distant blazars from being absorbed by the soft EBL photons under given conditions, see e.g.\ Refs.~\cite{Angelis2007,
Mirizzi2009, Dominguez2011b, Angelis2013, Meyer2013,
Troitsky2016}. As a consequence, the detection of hard TeV
photons from high red-shift sources by Imaging Atmospheric
Cherenkov Telescopes (IACTs) could be reasonably
interpreted even if the EBL intensity in NIR band is as high
as that measured by CIBER. This is recently verified by Kohri $et\ al$.\ \cite{Kori2017} on two hard spectra of TeV blazar 1ES1101-232 and H2356-309. In homogeneous external  magnetic fields, they assume the emitted TeV
photons can convert to ALPs in blazar jet and then this ALPs can re-convert into the photons inside the Milky Way (MW) Galaxy.

We aim to probe whether the excess radiation of CIBER is a new EBL component. So, we propose a method: thirteen observed spectra of distant TeV BL Lac objects are fitted by four theoretical spectra which involve theoretical EBL (Gilmore et al), Gilmore's EBL model with CIBER excess and Gilmore's EBL with/witout CIBER excess including photon/ALP coupling respectively; If the model including ALP and CIBER data has more significant advantage on explaining the $\gamma$-ray observations than that only with CIBER data, the tension can be alleviated or solved; Furthermore, if it is more compatible with the observations than theoretical EBL only model, then the excess component could be EBL in the presence of ALP; Otherwise the excess is less likely to be a new EBL component.

 If the excess is contributed by foreground yet undefined component \cite{CIBER2017} and the Gilmore's EBL model is more preferred by the observed spectra than that including ALP, there is no real discrepancy among the observations and the hypothetical ALP is not required. Therefore, investigating the origin of the excess radiation is necessary and significant.

The structure of the paper is as follows. In
Sec.~\ref{sec:ALP}, we briefly describe the ALP/photon
oscillation model. In Sec.~\ref{sec:magnetic}, we present
the magnetic fields configuration and ALP/photon conversion scenarios. In Sec.~\ref{sec:analyze},
we analyze the Fermi-LAT and IACTs data of 10 BL Lacs (13
spectra). In Sec.~\ref{sec:results}, the results of the spectral fits performed with the three models above are presented and discussed. Summary and conclusion are given in
Sec.~\ref{sec:conclusion}.

\section{ALP (axion-like particle)-photon oscillation}
\label{sec:ALP}
The ALP/photon oscillation are assumed to take place in the
magnetic field regions along the path of the propagating gamma-ray photons from the source to the earth.

We denote the y-axis direction as that of the propagating gamma-ray photons. We consider a photon/ALP beam of energy $E_{\gamma}$ propagating along the line of sight in a cold magnetized plasma. The beam for unpolarized photons is described with the density matrix $
\rho(y)=\begin{pmatrix}
A_{x}(y)& A_{z}(y)& a(y)
\end{pmatrix}^{\rm T} \bigotimes \begin{pmatrix}
A_{x}(y)& A_{z}(y)& a(y)
\end{pmatrix}^{\ast }$, where $A_{x}(y)$ and $A_{z}$
describe the photon linear polarization amplitudes along the
x and z axis, respectively, a(y) is the ALP amplitude. From
the Lagrangian of photon-ALP system, we can derive the beam
propagation equation for ultra-relativistic ALPs
 \cite{Raffelt1988, Tavecchio2012, Angelis2013, Masaki2017}
\begin{equation}\label{1}
i\frac{\mathrm{d\rho\left ( y \right ) } }{\mathrm{d}
y}=[\rho(y) ,M].
\end{equation}
Here, M denotes the photon-ALP mixing matrix, including the
mixing, refractive and the photon absorption effects. It is
defined as \cite{zhang2018}
\begin{equation}\label{2}
M=V^{\dagger }M_{\mathbf{B}_{\rm T}\parallel\mathbf{e}_{z}}V
\end{equation}
where $\mathbf{B}_{\rm T}$ is the component of the magnetic
field strength in the x-z plane,
$M_{\mathbf{B}_{\rm T}\parallel\mathbf{e}_{z}}$ represents the
mixing matrix for $\mathbf{B}_{\rm T}\parallel\mathbf{e}_{z}$
and V is the rotation matrix in the x-z plane. When
$\mathbf{B}_{\rm T}$ forms an angle $\psi$ with the z axis, the
matrix V is
\begin{equation}\label{3}
V=\begin{pmatrix}
\cos\psi  & \sin\psi & 0\\
-\sin\psi  & \cos\psi  & 0\\
 0 &  0 &1
\end{pmatrix}.
\end{equation}
 The Faraday effect is totally irrelevant at
 the energies (HE and VHE) considered in this paper and can be neglected, then the
 mixing matrix $M_{\mathbf{B}_{\rm T}\parallel\mathbf{e}_{z}}$
 can be written as
\begin{equation}\label{4}
M_{\mathbf{B}_{\rm T}\parallel\mathbf{e}_{z}}=\begin{pmatrix}
\Delta _{\perp } & 0 &0 \\
 0& \Delta _{\left | \right | } & \Delta _{a\gamma  }\\
 0& \Delta _{a\gamma  } & \Delta _{aa}
\end{pmatrix}.
\end{equation}

We denote the mean free path of $\gamma$-ray photon in
extragalactic space as $ \lambda_{\gamma}$, and the ALP
mass as $m_a$.

Then the various $\Delta$ terms in $M_{\mathbf{B}_{\rm T}
\parallel\mathbf{e}_{z}}$
are~\cite{Raffelt1988} $\Delta_\perp
= \Delta_{\rm pl} + 2\Delta_{\rm QED} +
\Delta_{\rm CMB}+i/2\lambda_{\gamma},$ $ \Delta_\parallel =
\Delta_{\rm pl}
+ (7/2)\Delta_{\rm QED} + \Delta_{\rm CMB}+i/2\lambda_{\gamma},
$
$\Delta_{a\gamma} = {g_{a\gamma} B_{\rm T}}/{2} $ and
$\Delta_{aa} = - {m_a^2}/{2 E_{\gamma} }$, where $g_{a\gamma}$ is the
coupling constant. The plasma effect $\Delta_{\rm pl}$ is
stemmed from an effective photon mass produced by charge
screening in plasma. $\Delta _{a\gamma}$ is photon/ALP
mixing term. The term $\Delta_{\rm QED}$ describes the photon one-loop
vacuum polarization. Photon-photon dispersion provided by
the cosmic microwave background (CMB) leads to the term
$\Delta_{\rm CMB}$ with a factor $(1+z)^{3}$ to account for the evolution of CMB density at red-shift
$z$, which is considered only in extragalactic space. The
Cotton-Mouton effect, for the energies and magnetic fields
considered here, can be reasonably neglected.

For the relevant parameters, numerically we find under the natural
Lorentz-Heaviside units \cite{Montanino2017}
\begin{eqnarray}
\Delta_{a\gamma}&\simeq &   1.5\times10^{-3}
\left(\frac{g_{a\gamma}}{10^{-12}\,\textrm{\rm GeV}^{-1}} \right)
\left(\frac{B_{\rm T}}{10^{-9}\,\rm G}\right) {\rm Mpc}^{-1}
\nonumber\,,\\
\Delta_{aa} &\simeq &
 -8 \times 10^{-2} \left(\frac{m_a}{
        {\rm neV}}\right)^2 \left(\frac{E_{\gamma}}{{\rm TeV}}
        \right)^{-1}
        {\rm Mpc}^{-1}
\nonumber\,,\\
\Delta_{\rm pl}&\simeq &
  -1.1\times10^{-7}\left(\frac{E_{\gamma}}{{\rm TeV}}\right)^{-1}
         \left(\frac{n_{\rm e}}{10^{-3} \,{\rm cm}^{-3}}\right)
         {\rm Mpc}^{-1}
\nonumber\,,\\
\Delta_{\rm QED}&\simeq &
4\times10^{-9}\left(\frac{E_{\gamma}}{{\rm TeV}}\right)
\left(\frac{B_{\rm T}}{10^{-9}\,\rm G}\right)^2 {\rm Mpc}^{-1}
\nonumber\,,\\
\Delta_{\rm{CMB}}&\simeq & 0.80 \times 10^{-1}
\left(\frac{E_{\gamma}}{{\rm TeV}}
\right)(1+z)^{3}~{\rm Mpc}^{-1} \,\ ,
\label{5}\end{eqnarray}
where we have used the relation of unit conversion:
$1\,\rm eV\simeq1.57\cdot10^{29}\,{\rm Mpc}^{-1}$, $1\,\rm g \simeq 5.60 \cdot
10^{32}~\rm eV$
 and $1\,\rm G \simeq 1.95 \cdot10^{-2}\,{\rm eV}^{2}$ \cite{Anchordoqui2018}.

 We split the beam path, in every mixing region, into
 many small cells with a homogeneous magnetic
 field in each cell. For an initial photon matrix
 $\rho_{\rm in}(y_0)=1/2 ~ {\rm diag}
(1, 1, 0)$ at $y_0$, the density matrix at position $y$ is
the solution of Eq.~\ref{2}, which have been given in the
literature, e.g.\ Refs.~\cite{Tavecchio2012, Angelis2013,
Montanino2017}, as
\begin{eqnarray}\label{6}
\rho(y)=U(y,y_{0})\rho(y_0) U^\dagger(y,y_{0}),
\end{eqnarray}
where $U(y,y_{0})=\prod^{N}_{0}U(y_{n+1},y_{n})$ is the transfer matrix. Then the photon
survival probability at the final polarization states
$\rho_{x}(y)=1/2 ~ {\rm diag}(1, 0, 0)$ and $\rho_{z}(y)=1/2
~ {\rm diag}(0, 1, 0)$ after propagating through a mixing
region is given by
\begin{eqnarray}\label{7}
P_{\gamma\rightarrow\gamma}(y,y_{0})=\rm Tr((\rho_{x}(y_0)+\rho_{z}(y_0))\rho(y)).
\end{eqnarray}.

If we neglect the absorption and the magnetic field strength
$B_{\rm T}$ is the same constant in all the cell but with a
random angle $\psi$ in each one, the condition for
significant conversion is
\begin{eqnarray}\label{condition}
g_{11}^{2}\,(B_{\rm T\mu G})^{2}\,r_{\rm kpc}\,l_{\rm kpc} \geq 2900,
\end{eqnarray}
where $r$ is the distance the beam have propagated, and $l$
is the cell length \cite{Meyer2014}. We use the
notation above $g_{\rm X}=g_{a\gamma}/(10^{\rm -X}~\rm GeV^{-1})$ and
$A_{\rm X}=A/\rm X$, where A represents $B_{\rm T}$, $r$, or $l$.

 In a simple case where the magnetic field is homogeneous and the resonant conversion \cite{Hochmuth2007} is neglected, the conversion probability of a photon to an ALP is \cite{Kori2017, Hooper2007}
\begin{equation}\label{probability}
P_{\gamma\rightarrow a}=\frac{1}{2[1+(\frac{E_{\rm crit}}{E_{\gamma}})^{2}]}\rm sin^{2}[\frac{g_{a \gamma}B_{\rm T}r}{2}\sqrt{1+(\frac{E_{\rm crit}}{E_{\gamma}})^{2}}\ ]
\end{equation}
where $E_{\rm crit}$ is a critical energy, see
e.g.\ Refs.~\cite{Angelis2008},
\begin{equation}\label{Ecrit}
E_{\rm crit}\simeq625(\frac{m_{a}}{5\cdot10^{-10}\rm eV})^{2}(\frac{10^{-9}\rm G}{B_{\rm T}})(\frac{
10^{-11}{\rm GeV}^{-1}}{g_{a\gamma }})\,\rm  GeV
\end{equation}
From Eq.~\ref{probability}, $P_{\gamma\rightarrow a}$ become energy-independent and sizable when $E_{\gamma}>E_{\rm crit}$ and $g_{a \gamma}B_{\rm T}r/2>1$.

\section{magnetic field configuration and conversion scenarios}
\label{sec:magnetic}
\subsection{Magnetic field configuration}
\begin{figure}[!h]
\centering
  \begin{minipage}[b]{0.49\textwidth}
    \centering
    \includegraphics[width=8cm]{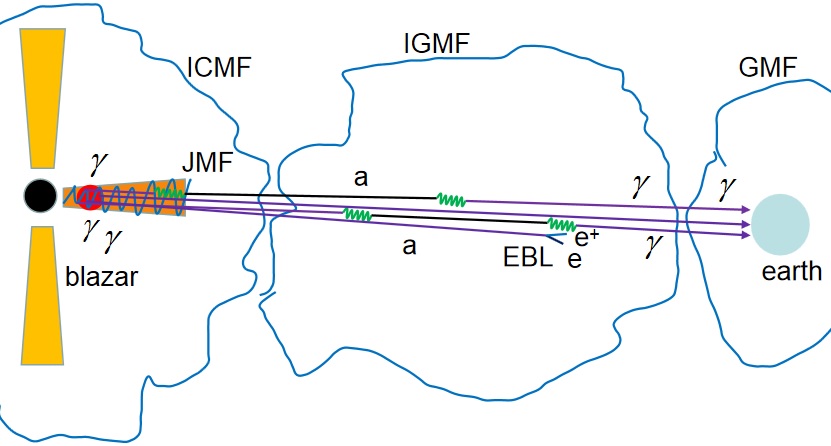}
  \end{minipage}\\
  \begin{minipage}[!h]{0.49\textwidth}
    \centering
    \includegraphics[width=8cm]{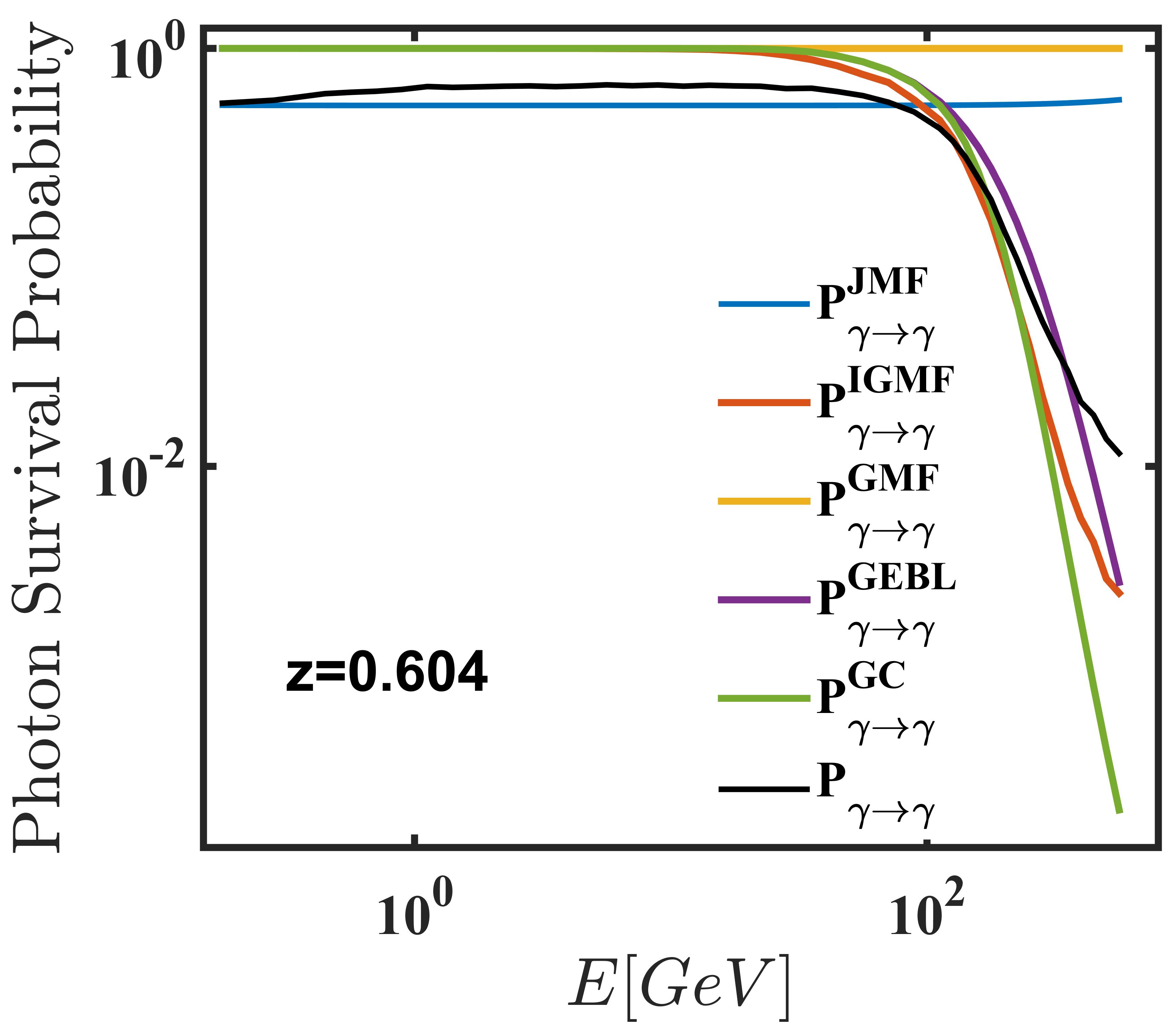}
  \end{minipage}
  \caption{\small{Top panel: Cartoon of the formalism
  adopted in this article, where the photon/ALP beams
  propagate from the VHE emission region (red solid circle)
  to the Earth and pass through four regions of magnetic fields: JMF, ICMF, IGMF and GMF
  respectively. The symbols {\it a} means ALP and the green
  crooked line represent ALP $\Longleftrightarrow$ photon
  conversion. Bottom panel: The survival probability for the
  $\gamma$-ray photon from PKS 1424+240. The meaning of each
  curves are given in the graph. The probability $P_{\gamma\rightarrow\gamma}
  ^{\rm GEBL}$ is the absorption function for Gilmore EBL and $P_{\gamma\rightarrow\gamma}^{\rm GC}$ represent the attenuation by Gilmore EBL and CIBER EBL. $P_{\gamma\rightarrow\gamma}^{\rm GC}$ is the total probability. The values of ALP parameters are $m_a=0.9\,\rm neV$ and $g_{12}=2.9$ for which only the conversion in jet and in IGMF is significant.}}
  \label{fig:2}
\end{figure}

Along the entire path of the ALP/photon beam from the source to the Earth, the magnetic fields are ubiquitous. The environment
along the line of sight can provide some cues about the
magnetic fields. It is widely believed that BL Lacs reside
in elliptical galaxies embedded in small galaxy clusters or
galaxy groups \cite{Farina2015}. Thus the photon/ALP beam
crosses altogether five different regions of plasma and
magnetic field configurations. As done in \ Refs.~\cite{Meyer2014},
we ignore the mixing in the magnetic field of the host galaxy due
to the small conversion probability in this region. Hence we
only need to consider the environments of four regions the $\gamma$-rays photons pass through: 1. the jet magnetic field (JMF); 2. intra-cluster magnetic field (ICMF); 3. the intergalactic magnetic fields (IGMF); 4. the galactic magnetic field (GMF), as show in Fig.~\ref{fig:2}.

The ALP/photon conversion, with allowed ALP parameters, in IGMF has recently been found to be significant and could enhance the spectral hardening for cosmic TeV photons, if the IGMF obtained from large-scale cosmological simulations were adopted \cite{Montanino2017}. We will apply the simulated IGMF to model the observations for the first time. At the same time, a realistic magnetic fields configurations of the ALP/photon mixing region in blazar jet will be considered.

 In the following subsections, we discuss the observational evidence and model assumptions for each environment respectively.

 \emph{JMF}---According to the observations and theory
 diagnostics, the magnetic field in BL Lac jet appears to be  large
 scale coherent field ranging from 0.1 pc near the central
 engine up to kpc scales along the jet, this field is mainly
 ordered and the component traversed to the jet axis is
 predominant \cite{Pudritz2012, Tavecchio2015}. Based on
 this observational evidence and theoretical deduction, we
 apply the model proposed by Tavecchio et al.\ \cite{Tavecchio2015} or Mena et al.\ \cite{Mena2013}, in which $B_{\rm T}(y)\propto\frac{1}{y}$
 and y is the coordinate along the jet axis.

In this model, the plasma effect can be neglected according to its numerical value in Eq.~\ref{5}, given the typical relevant-parameter values (for example, the electron density $n_{\rm e} \sim 5\cdot10^{4}\,\rm cm^{-3}, \langle B_{\rm T}\rangle \sim 0.1-1\,\rm G$. Thus, there are four free parameters in the mixing model to be determined: the distance $d_{\rm VHE}$ of VHE emitting region from the central engine, the magnetic field strength $B_{\rm VHE}=B_{T}(d_{\rm VHE})$ in VHE emitting region, the jet Doppler factor $\delta$, and
the length $L$ of the large scale coherent magnetic field.
We adopt the typical values $L \sim 1\,\rm kpc$ for every
source. $B_{\rm VHE}$, $\delta$ and the size $R_{\rm VHE}$
of emission region can be determined, in principle, by fitting to the multi-frequency observations with theoretical models for blazar emission such as synchrotron self Compton (SSC) \cite{Xue2016}. But for simplicity we adopt a single central value for all sources, and they are $B_{\rm VHE}$=0.2\,G, $\delta$=30 and $R_{\rm VHE}$=$2\times10^{16}$\,cm. Thus $d_{\rm VHE}$ can be estimated by $d_{\rm VHE}\approx\delta R_{\rm VHE}/2$. An efficient conversion can realize, even if the coupling constant $g_{a\gamma}$ is down to $10^{-12}~{\rm GeV}^{-1}$ for these values of $B_{\rm T}$ and L according to $g_{a \gamma}B_{\rm T}r/2>1$.
Finally, we calculate the photon survival probability
$P_{\gamma \rightarrow \gamma }^{\rm jet}$ in this region, as show in
Fig.~\ref{fig:2} (for the source PKS 1424+240).

\emph{ICMF}---Faraday rotation measurements (FMs) and
synchrotron emission at radio frequencies well establish
that turbulent magnetic field with strength a few $\mu \rm G$
exists in the center of poor clusters or groups in which BL
Lacs may be harboured \cite{Feretti2012}. The turbulent
ICMF can be modeled with a divergence-free (i.e.,~$\nabla
\cdot \mathbf{B}$=0) homogeneous and isotropic Gaussian field,
which is described in detail in\ Refs.~\cite{Meyer2014}. In
this ICMF model \cite{Fermi2016}, there are five complete
free parameters, if considering the degeneracy between the
magnetic field strength $B_{0}$ at the cluster centre and
the index $\eta$. The mixing probability is sensitive to
three parameters, i.e.,~$B_{0}$, the cluster radius
$r_{\rm max}$ and the correlation length $l$ that is related to
the free parameters of the turbulence spectrum.
\begin{figure}[!h]
\centering
  \begin{minipage}[b]{0.45\textwidth}
    \centering
    \includegraphics[width=7.0cm]{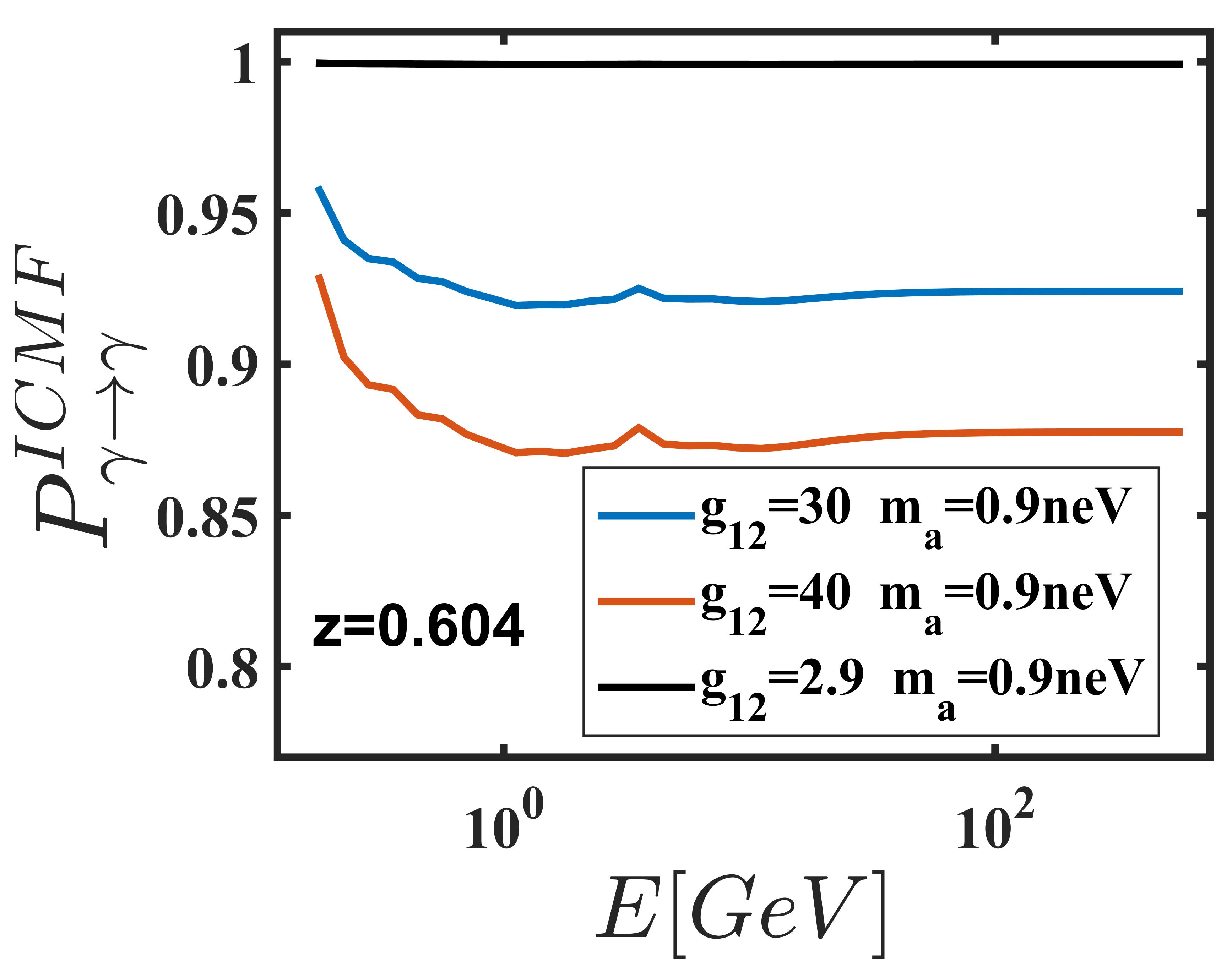}
  \end{minipage}\\
  \begin{minipage}[!h]{0.45\textwidth}
    \centering
    \includegraphics[width=7.0cm]{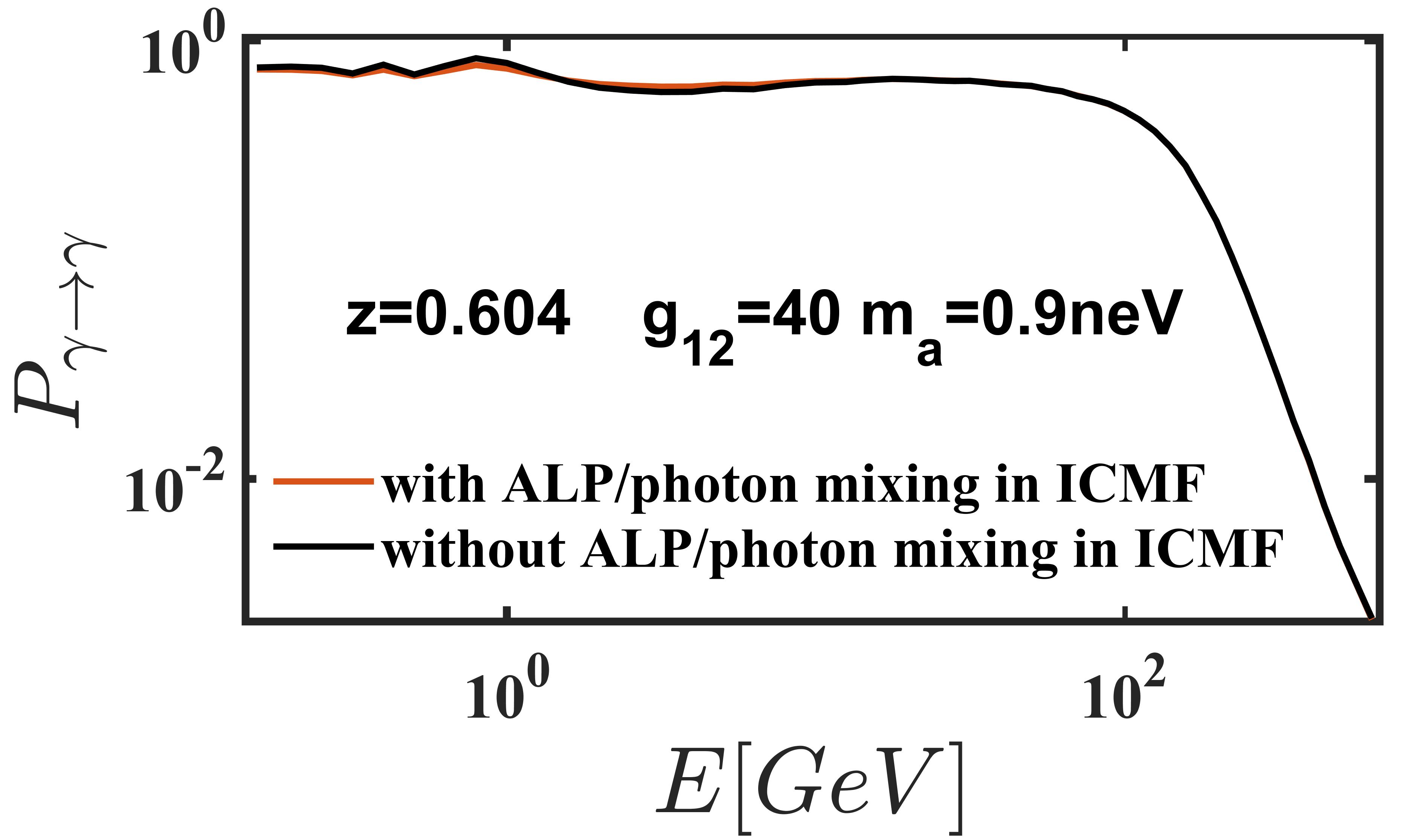}
  \end{minipage}
  \begin{minipage}[!h]{0.45\textwidth}
    \centering
    \includegraphics[width=7.0cm]{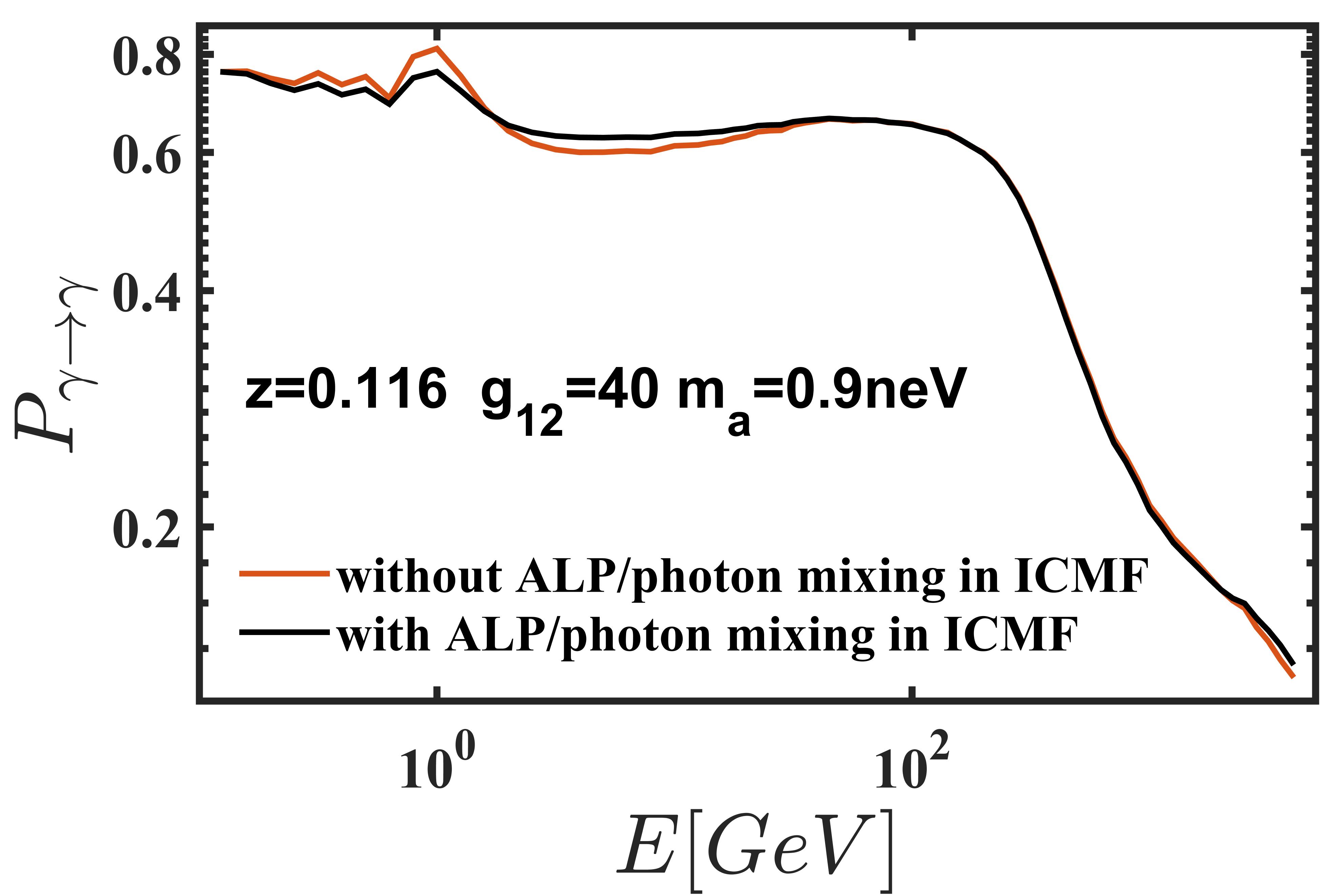}
  \end{minipage}
  \caption{\small{Top panel: the photon survival probability in the intra-cluster magnetic field $P_{\gamma\rightarrow\gamma}^{\rm ICMF}$
 varies with the photon-ALP coupling constant $g_{a\gamma}$ for PKS 1424+240. Middle and Bottom panel: the effect of $P_{\gamma\rightarrow\gamma}^{\rm ICMF}$ on the total probabilities $P_{\gamma\rightarrow\gamma}$ (including CIBER EBL absorption) for PKS 1424+240 and PKS 2155-304 respectively. The coherence length of IGMF here is determined by the turbulent spectrum, that is $l$=3.05\,kpc. The values of other cluster-related parameters refer to Table 1 in \ Ref.~\cite{Meyer2014}: the maximum turbulence scale $k_{\rm H}$=3.14\,${\rm kpc}^{-1}$, the minimum turbulence scale $k_{\rm L}$=0.18\,${\rm kpc}^{-1}$, the turbulent spectrum index q=-11/3, the electron number density $n_{0}$=0.001\,$\rm cm^{-3}$, the power-law index for $B(r)$, $\eta\beta$=2/3, the core radius of cluster $r_{\rm core}$=100\,kpc, $B_{0}$=1\,$\mu {\rm G}$, $r_{\rm max}$=300\,kpc.}}
  \label{fig:3}
\end{figure}

Though the source PKS 1424+240 in our sample may reside
in an intermediate cluster \cite{Rovero2016}, its ICMF
remains poorly known and the relevant parameters cannot be
determined by insufficient information about the cluster environment. Therefore, the fiducial, corresponding to a small cluster, model parameters of Table 1 in~Ref.~\cite{Meyer2014} are adopted for all the sources in our
sample except $l$. Here, we self-consistently take $l=3.05$\,kpc that is resulted from the calculation with the turbulence spectrum \cite{Meyer2014}. Then, the conversion probability $P_{\gamma \rightarrow \gamma }^{\rm ICMF}$ is small
and even negligible for $g_{12}\leq40$. This
can be roughly proved with Eq.~\ref{condition}. Since the radius $r_{\rm max}\sim 300$\,kpc for a typical value, the average transversal components $\left \langle B_{\rm T}\right \rangle$ in the whole region is
about $0.25$\,$\mu$G for $B_{0}=1$\,$\mu$G, and the coupling
constants $g_{12}\leq40$, these
parameter values lead to $g_{11}^{2} \left \langle B_{\rm T_{\mu \rm G} }\right \rangle^{2} (r_{\rm max})_{\rm kpc}$ $l_{\rm kpc}$ $<2900$.
This can also be verified through numerical calculation and is  shown in Fig.~\ref{fig:3}. The photons converted to the ALPs are less than 10\% when $g_{12}<35$, and the
affection of ALP/phothon mixing in this region on the photon
survival probability over the entire path
$P_{\gamma\rightarrow\gamma}$ is smaller. Even if considering the value of the most significant mixing probability $g_{12}=40$,
the deviation in $P_{\gamma\rightarrow\gamma}$ caused by
omitting the ALP/photon mixing in this region is less than
5\% in VHE band. So in order to reduce the number of
free parameters, we neglect the conversion in this
region for $g_{12}\leq40$.

\emph{IGMF}---FMs of polarized extragalactic
sources~\cite{Blasi1999} and CMB observations~\cite{Planck2016} give the lowest upper limits on IGMF
on the largest cosmological scales \cite{Montanino2017},
i.e.~the average field strength of the IGMF with a
coherence length $l_{\rm c} \sim {\mathcal O}$ (1\,Mpc) is $\lesssim
{\mathcal O}$ (1\,nG) \cite{Pshirkov2016}. According to Eq.~\ref{Ecrit} and the present constraints on the
ALP-parameter space from other observations \cite{Montanino2017}, the conversion in the IGMF with a cell-like structure \cite{Angelis2013} is not important to reduce the cosmic opacity. However, a more realistic treatment for the IGMF have been proposed by Montanino et al.\ \cite{Montanino2017}, i.e.,~the IGMF is obtained from
large-scale cosmological simulations produced with the MHD
code ENZO \cite{Bryan2014}. Such simulated IGMF ($B_{\rm T}$) would enhance to $10^{-7}$\,G in the large scale filaments of cosmic matter
so that an efficient conversion could be achieved for ALP parameters that are not excluded by observations \cite{Montanino2017}.

In this work, we use this type simulated IGMF, and 98 column data of $B$ along the random line of sights up to z=1 are obtained from the web side of Vazza \cite{Vazzaweb}. These line of sights are extracted from a $200^{3}$\,${\rm Mpc}^{3}$ (comoving) volume simulated with $2400^{3}$ cells and dark matter particles (for a co-moving resolution of 83.3 kpc/cell). In their cosmological simulation, the magnetic field $B$ is initialized to $B_{0}$=1\,nG (comoving) at $z$=50, and is assumed to be uniform in all directions. Thus, we treat the IGMF as a turbulent field as in ``cell" model, but the size of each cell or domain equal to the simulation resolution of
 83.3\,kpc/cell.

 We calculate the mean free path of $\gamma$-ray photon in the nth-cell by $\lambda _{\gamma}^{(n)}=l/\tau_{\rm G/GC}^{(n)}$,
 where $\tau_{\rm G/GC}^{(n)}$ is the optical depth of Glmore
 EBL (see Fig.~\ref{fig:1}) and its combination with
 CIBER data (Fig.~\ref{fig:1}), respectively. We take the lower bound of
 the data with systematic error and one $\sigma$ statistical
 error in each bin, see Fig.~\ref{fig:1}. Even so, this
 lowest IR intensity of CIBER measurements can be in tension
 with the constraints inferred by $\gamma$-ray observations \cite{Kori2017}.
 Fig.~\ref{fig:2} show photon survival probabilities
 corresponding to the combined CIBER and Gilmore EBL(green
 curve), the Gilmore EBL (purple curve). The red-shift
 evolutions of the CIBER IR background due to evolutions of
 sources are calculated, assuming the evolution factor
 $f_{\rm evol}=1.7$, with the method in \ Ref.~\cite{Biteau2015} which is effective until $z \sim 0.6$. Note that the most high-red-shift source PKS1424+240 in our sample just has a
 red-shift $z=0.604$.

The QED vacuum polarization terms and plasma effect terms
are small compared to the other ones, hence they are not
considered. At the highest energy band, the effect of CMB
photon scattering is important \cite{Dobrynina2015,
Montanino2017} and must be considered. We have 98 random
magnetic field samples run along the line of sights extracted
from large-scale cosmological simulation so that we
can simulate 98 random realizations of the turbulent field.

The photon survival probability $P_{\gamma \rightarrow
\gamma }^{\rm IGMF}$ in this region is calculated after carrying
out 98 Monte Carlo simulations, see Fig.~\ref{fig:2}, for
$z$=0.604. In order to obtain the total probability
$P_{\gamma\rightarrow \gamma}$ below, we need to calculate
the probability the photon converse to the ALP with state
$\rho_{a}=\rm diag(0,0,1)$, $P_{\gamma \rightarrow a
}^{\rm IGMF}=\rm Tr(\rho_{a}\rho)$,
and that ALP converse to ALP, $P_{a \rightarrow
a}^{\rm IGMF}=\rm Tr(\rho_{a}U\rho_{a}U\dag)$, in advance.

\emph{GMF}---We model the galactic magnetic field $B$ with the regular component of
magnetic field presented by Jansson and Farrar \cite{Jansson2012}. The
turbulent component is not considered here, as the typical
coherence length is far smaller than the photon-ALP
oscillation length. Neglecting the plasma effect, we
calculate the photon survival probability $P_{\gamma
\rightarrow \gamma }^{\rm GMF}$ under this magnetic-field
structure. Fig.~\ref{fig:2} shows the curve of $P_{\gamma
\rightarrow \gamma }^{\rm GMF}$ for $g_{12}=2.9$ and
the conversion is puny in this condition. For a significant conversion in this region, we could roughly constraint the coupling constant $g_{a\gamma}$ determined from $g_{a \gamma}\frac{B_{\rm T}}{\mu G}\frac{r}{\rm 20kpc}\gtrsim3\times10^{-11}\,{\rm GeV}^{-1}$, that is $g_{a\gamma}\gtrsim3\times10^{-11}\,{\rm GeV}^{-1}$.

The photon survival probability on the whole path from
the source to detector is
\begin{equation}\label{P}
\begin{aligned}
P_{\gamma\rightarrow\gamma}&=P_{\gamma \rightarrow \gamma
}^{\rm JMF}P_{\gamma \rightarrow \gamma }^{\rm ICMF}(P_{\gamma
\rightarrow \gamma }^{\rm IGMF}P_{\gamma \rightarrow \gamma
}^{\rm GMF}+P_{\gamma \rightarrow a }^{\rm IGMF}P_{a \rightarrow
\gamma }^{\rm GMF})\\
 &+P_{\gamma \rightarrow \gamma }^{\rm JMF}P_{\gamma \rightarrow
 a }^{\rm ICMF}(P_{a \rightarrow \gamma }^{\rm IGMF}P_{\gamma
 \rightarrow \gamma }^{\rm GMF}+P_{a \rightarrow a }^{\rm IGMF}P_{a
 \rightarrow \gamma }^{\rm GMF})\\
 &+P_{\gamma \rightarrow a }^{\rm JMF}P_{a \rightarrow \gamma
 }^{\rm ICMF}(P_{\gamma \rightarrow \gamma }^{\rm IGMF}P_{\gamma
 \rightarrow \gamma }^{\rm GMF}+P_{\gamma \rightarrow a
 }^{\rm IGMF}P_{a \rightarrow \gamma }^{\rm GMF})\\
 &+P_{\gamma \rightarrow a }^{\rm JMF}P_{a \rightarrow a
 }^{\rm ICMF}(P_{a \rightarrow \gamma }^{\rm IGMF}P_{\gamma
 \rightarrow \gamma }^{\rm GMF}+P_{a \rightarrow a }^{\rm IGMF}P_{a
 \rightarrow \gamma }^{\rm GMF}),\\
 \end{aligned}
\end{equation}
\\
 where $P_{a \rightarrow \gamma }^{\rm IGMF}=2P_{\gamma
\rightarrow a }^{\rm IGMF}$, $P_{a \rightarrow a }^{\rm ICMF}=1-P_{a
\rightarrow \gamma }^{\rm ICMF}$, $P_{a \rightarrow \gamma
}^{\rm ICMF}=2(1-P_{\gamma \rightarrow \gamma }^{\rm ICMF})$, $P_{a
\rightarrow \gamma }^{\rm GMF}=2(1-P_{\gamma \rightarrow \gamma
}^{\rm GMF})$, and $P_{\gamma \rightarrow a }^{\rm GMF}=1-P_{a
\rightarrow \gamma }^{\rm GMF}$. When neglecting the cluster
mixing, we have $P_{\gamma \rightarrow \gamma }^{\rm ICMF}=1$.

Fig.~\ref{fig:2} reveals the conversion in jet dominates
$P_{\gamma\rightarrow\gamma}$ at HE, whereas the conversion
in jet and in extragalactic media together contributes to the total conversion at VHE for $m_a=0.9\,\rm neV$ and $g_{12}=2.9$. The critical energy $E_{\rm crit}$ of ALP/photon oscillation in jet is at HE due to the large GMF, whereas that in IGMF is at VHE. Thus the ALPs converted from photons at HE cannot convert back to photons so that abundant photons at HE lost, and the net effect is a hardening in the observed HE-VHE spectrum.
\subsection{Conversion scenarios}
Broadly speaking, two complementary ALP/photon conversion scenarios for enhancing transparency of the VHE universe have been proposed \cite{Troitsky2016, Horn2012, Montanino2017}:~(a) photons convert to ALPs in the gamma-ray sources and then  back-conversion happens in GMF, e.g.\ Ref.~\cite{Simet2008}, (b) the photon/ALP oscillations take place in IGMF, e.g. Ref.~\cite{Angelis2007, Angelis2013}. We will mainly investigate the photon/ALP oscillations which in turn take place in JMF, IGMF, and GMF along the way of the ALP/photon beam. In this scenario, the ALP mass is limited to $m_a<2.5\,\rm neV$ for an efficient conversion with $E_{\rm ctit}<1\,\rm TeV$ and $g_{12}\sim10$ in IGMF. We take the lower limit of $m_a$ as 0.1\,neV, which is small enough and no essential effect would be on the conversion probability for smaller mass. Due to the observation constraints on the ALP parameter space \cite{Montanino2017}, the coupling constant is limited to $g_{12}<5$ for $m_{a}\in[0.1\,\rm neV,2.5\,\rm neV]$. Nevertheless, we consider the space of coupling constant $g_{12}\in[2.9,40]$, for which the conversion in ICMF can be neglected. This parameter space $(m_{a},g_{12})\in[0.1\,\rm neV,2.5\,\rm neV]\times[2.9,40]$ have not been excluded by CAST \cite{CAST2017} and can be
probed by the future IAXO experiment.

For these ALP parameters, substantial photons at HE and VHE can convert to ALPs in JMF, and it always means a reduction of the photon flux, although a part of these ALPs may convert back to photons in IGMF or GMF. The photons or ALPs at HE can reach MW galaxy almost without converting or damping in IGMF. Whereas the photons or ALPs above the critical energy ($\sim$500 GeV) would mix or be damped due to the $e^{\pm}$ conducing reaction during the propagation in IGMF: $\gamma\rightarrow\gamma$, $\gamma\rightarrow e^{\pm}$, $\gamma\rightarrow a$, $\gamma\rightarrow a\rightarrow \gamma$, $\cdot\cdot\cdot$ or $a\rightarrow a$, $a\rightarrow \gamma$, $a\rightarrow \gamma\rightarrow e^{\pm}$, $a\rightarrow \gamma\rightarrow a$, $\cdot\cdot\cdot$. We can show that the probability $P_{a\rightarrow\gamma}^{\rm IGMF}$ is sensitive to the coupling constant $g_{a \gamma}$ and the distance of propagation. When $g_{a \gamma}$ is not large enough (e.g.,~$g_{12}<40$, z$\sim$0.3), $P_{a\rightarrow\gamma}^{\rm IGMF}$ increases with $g_{a\gamma}$. But when $g_{a\gamma}$ is large enough, $P_{a\rightarrow\gamma}^{\rm IGMF}$ stays the same (saturation) or decreases. This is because the process of $a\rightarrow \gamma$ becomes more and more important as the $g_{a\gamma}$ get larger, and the process of $a\rightarrow \gamma\rightarrow a$ would become significant when $g_{a\gamma}$ is large enough. The case for $P_{\gamma\rightarrow\gamma}^{\rm IGMF}$ is similar, namely that $P_{\gamma\rightarrow\gamma}^{\rm IGMF}$ increases with $g_{a\gamma}$ and then becomes stable when $g_{a\gamma}$ gets large enough due to the transition of the dominant process from $\gamma\rightarrow a$ and $\gamma\rightarrow a\rightarrow \gamma$ to $\gamma\rightarrow a\rightarrow \gamma\rightarrow a$.

  If $g_{12}<20$ the conversion in GMF is insignificant, while, for $g_{12}>30$, the VHE ALP/photon oscillation in IGMF reach or is close to saturation after a distant propagation (e.g.,~$z>0.2$) so that the conversion for VHE photons in GMF is subordinate to that in IGMF. Therefore, for a rough estimate we neglect the conversion in the GMF and the survival probability $P_{\gamma\rightarrow\gamma}$ for VHE photons could be simplified as
  \begin{equation}\label{simP}
P_{\gamma\rightarrow\gamma}=P_{\gamma\rightarrow\gamma}^{\rm JMF}
  P_{\gamma\rightarrow\gamma}^{\rm IGMF}+P_{\gamma\rightarrow a}^{\rm JMF}P_{a\rightarrow\gamma}^{\rm IGMF}.
\end{equation}
     The variation of function $P_{\gamma\rightarrow\gamma}$ at VHE is dominated by $P_{\gamma\rightarrow\gamma}^{\rm IGMF}$ and $P_{a\rightarrow\gamma}^{\rm IGMF}$, since the conversion in JMF is always sufficient and $P_{a\rightarrow\gamma}^{\rm JMF}$ is relatively stable. Therefore, $P_{\gamma\rightarrow\gamma}$ increases with $g_{a\gamma}$ and then reach an extremum when $g_{a\gamma}$ is large enough. For a higher z and $E$ considered, the $g_{a\gamma}$ corresponding to the extremum may be required to be lower. Moreover, for our samples with z$>0.1$, we could take most of the larger values of $P_{\gamma\rightarrow\gamma}$ (the values of $P_{\gamma\rightarrow\gamma}$ for $g_{12}<66$) although we consider only the ALP parameters space $(m_{a},g_{12})\in[0.1\,\rm neV,2.5\,neV]\times[2.9,40]$.

   In a word, a part of emitted $\gamma$-rays photons convert to ALPs in JMF and then only some of these ALPs at HE convert back to photons in GMF; whereas the ALP/photon above the critical energy ($\sim$500 GeV) could continue to oscillate in IGMF, where the processes of $a\rightarrow \gamma$ and $\gamma\rightarrow a\rightarrow \gamma$ could allow the original TeV photon to escape absorption by the EBL, so that the photon flux observed toward the highest energy could be enhanced relative to that without ALP. This mechanism could shape a harder spectra. This conversion pattern is similar to the conversion scenario (b).

  We consider the complementary conversion scenario (a) now. If we consider larger ALP-parameter values rather than limiting to the region $(m_{a},g_{12})\in[0.1\,\rm neV,2.5\,neV]\times[2.9,40]$, e.g.,~allowing $m_{a}>2.5\,\rm neV$ and $g_{12}>40$, the conversion for TeV photons taking place in the source and in the Milky Way would be dominant due to inefficient oscillations in intergalactic space (the critical energy is too high). Furthermore, the resultant simulated IGMF could be weaker if the magnetic field is initialized to $B_{0}<1\,$nG at $z$=50 in the large-scale cosmological simulations \cite{Vazza2017}. In this case, the conversion pattern (a) could be dominant although the coupling constant reaches down to $g_{12}\sim10$ for $m_{a}\gtrsim1\,\rm neV$ and $B_{T}\thicksim1\,$nG. This scenario with a negligible conversion in IGMF will be discussed in Sec.~\ref{sec:results}.

\section{analyzing the Fermi-LAT and IACTS data}
\label{sec:analyze}
We build four absorption models about the total photon survival probability from the source to the earth. (i) it only involves Gilmore EBL (refer to as GEBL model below); (ii) it includes Gilmore EBL combined with CIBER ``excess" (refer to GC model below); (iii) it includes Gilmore EBL with/without CIBER ``excess", taking into account the ALP/photon oscillation (refer to GCA/GA model below). Each of the four models together with an assumed-intrinsic spectrum will be used to fit 13 observed BL Lac objects (BL Lacs) spectra at high energy (100 MeV $\leq$ E $\leq$ 300 GeV, HE) and VHE. In order to test CIBER EBL and ALP, we need to compare the goodness-of-fit of four theoretical spectra based on the four absorption models.
\begin{table*}[t]
\caption{BL Lac spectra at HE and VHE used in this paper and their best-fitting models of intrinsic spectrum.}
\begin{ruledtabular}
\begin{tabular}{lrrrrccccc}
   Source& Redshift & Experiment(VHE Obs. Period) & Energy(GeV)
   & lon/lat$[^{\circ}]$ &function\\
\hline
PKS 2155-304 &0.116 & Fermi-LAT+HESS(2008) \cite{HESS215509}& 0.28-3340 & 17.7/-52.3 & LP \\[3pt]
PKS 2155-304 &0.116 & Fermi-LAT+HESS(2013) \cite{Sanchez2015}& 0.15-3180 & 17.7/-52.3 & LP \\[3pt]
1ES 0229+200 &0.14 & Fermi-LAT+HESS(2005-2006) \cite{Vovk2012, HESS2007} & 8.59-11500 & 153.0/-36.6 & LP \\[3pt]
1ES 0229+200 & 0.14 & Fermi-LAT+VERITAS(2009-2012) \cite{Vovk2012, VERITAS2014}& 8.59-7640  & 153.0/-36.6 & PL\footnotemark[1] \\[3pt]
1ES 1218+304 & 0.182 & Fermi-LAT+VERITAS(2008-2009) \cite{Taylor2011, VERITAS1218}& 0.67-1870  & 186.2/-82.7 & PLC \\[3pt]
1ES 1101-232 & 0.186 & Fermi-LAT+HESS(2004-2005)\cite{Belikov2011, Aharonian2007} & 1.62-2940 &  273.2/33.1 & PL \\[3pt]
1ES 0347-121 & 0.188 & Fermi-LAT+HESS(2006) \cite{Tanaka2014, HESS0347}& 1.69-2910 & 201.9/-45.7 & PL  \\[3pt]
1ES 1011+496 & 0.212 & MAGIC(2014) \cite{MAGIC2016}&
79.4-3060 & 165.5/52.7 & PLC \\[3pt]
1ES 0414+009 & 0.287 & Fermi-LAT+HESS(2005-2009) \cite{HESS0414}& 0.17-1120 & 191.8/-33.2 & PL \\[3pt]
OT081 & 0.322 & Fermi-LAT+HESS(2016) \cite{Schussler2017}&
0.17-1020 & 34.9/17.7 & LP \\[3pt]
PKS 0477-439 & 0.343? & Fermi-LAT+HESS(2009-2010) \cite{Fermi2010}& 0.17-2444 & 248.8/-39.9 & LP \\[3pt]
PKS 1424+240 & 0.604 & Fermi-LAT+VERITAS(2009) \cite{VERITAS2014a}& 0.185-515 & 29.5/68.2 & LP \\[3pt]
PKS 1424+240 & 0.604 & Fermi-LAT+VERITAS(2013) \cite{VERITAS2014} & 0.241-530 & 29.5/68.2 & LP \\[3pt]
\end{tabular}
\end{ruledtabular}
\footnotetext[1]{For some ALP parameter values in the GCA model and all parameter values in the GA model, the best-fitting function is LP in the models with ALP. It is worth emphasizing that for the best-fitting ALP values ($n_a$=0.1\,neV and $g_{12}$=4.9) of the global fit of all the spectra in the GCA model the chosen function is PL.}
\label{tb1}
\end{table*}

\subsection{Spectra selection}
Blazars are a subclass of active galactic nuclei (AGN) and
their jet are directed toward us. When the broad emission
lines in their optical spectra is weak, this blazar is
called a BL Lac object, i.e.,~\cite{Urry1995}. BL Lacs has a
broad-band spectral energy
distribution (SED) and are the dominant extragalactic TeV
sources. So far about 57 TeV BL Lacs have been detected by
IACTs \cite{TeVCat}. In this study, we only consider the BL Lac spectra that satisfy the following criteria: 1. the highest energy beyond 800 GeV after the redshift  correction; 2. the redshift of sources is relatively certain and larger than 0.1. 3. the observations at HE and VHE are simultaneous, or the source did not exhibit obvious variability in the $\gamma$-ray regime. 4. have at least 2 data points at HE and 5 at VHE.

  Thirteen spectral measurements of 10 BL Lacs are chosen, which are widely distributed on a redshift scale of 0.1 to 0.61. Except for 1ES1011+496, all spectra combine the Fermi-LAT data at HE and the IACTs data at VHE. The data quality is high for 1ES1011+496 as the data at HE-VHE band is only from MAGIC. Given its high redshift and have sufficient spectral measurements around TeV, PKS 0477-439 is chosen despite the
fact that its redshift is not completely certain (z=0.343, $\geq$ 97\%) \cite{Muriel2015}. 1ES
0229+200, 1ES 1101-232, 1ES 0347-121 and PKS 1424+240, may be hard spectrum sources with an intrinsic photon index $\Gamma_{\rm
VHE}\sim1.5$ \cite{HESS2007, Aharonian2007, HESS0347121, Furniss2013}.
PKS 1424+240 is one of the most distant currently known VHE
$\gamma$-ray emitter and its red-shift was confirmed in 2017
\cite{Paiano2017}. In Table~\ref{tb1}, we summarize the
spectra including references. Except for 1ES 1101-232(2004) and 1ES 0229+200(2005), the Fermi-LAT data and IACTs data in
each spectrum were measured (quasi) contemporaneously. We note that no significant variation of the VHE $\gamma$-ray flux was found in 1ES 0229+200 during its observed period of 2004 and in 1ES 1101-232(2005) \cite{HESS2007, Aharonian2007}.

\subsection{Intrinsic spectra and method}
 \emph{Functions for the intrinsic spectrum}---There are usually three functions used to fit intrinsic spectrum of blazar: power-law
 (PL), log-parabola (LP), and power-law with exponential cut-off (PLC). The two-parameter PL spectrum $\phi_{\rm PL}=\phi_{0}(E/E_{0})^{-\alpha}$ where $\alpha$ is the photon spectral index and it is constrained by the particle acceleration theory to $\alpha\geq1.5$, $\phi_{0}$
 is the flux normalization, and $E_{0}$ is the reference
 energy. The LP spectrum have a nonzero spectral curvature
 $\phi_{\rm LP}=\phi_{0}(E/E_{0})^{-s-t\, {\rm log}(E/E_{0})}$
 with the additional curvature parameter $t>0$, and also
 $\langle s+t\, {\rm log}(E/E_{0})\rangle \geq1.5$. We
 describe the PLC spectrum with 3 free parameters
 $\phi_{\rm PLC}=\phi_{0}(E/E_{0})^{-\alpha}{\rm exp}(-E/E_{\rm cut})$, where $E_{\rm cut}$ is the cut-off
 energy. The theoretical observed spectra are predicted by
  functions of photon survival probability and the intrinsic
  spectra
 \begin{equation}\label{eq:model}
\phi_{\rm G/GC/GCA/GA,X}=P_{\rm G/GC/GCA/GA}\phi_{\rm X},
 \end{equation}
 where $P_{\rm G}$=exp$(-\tau_{\rm G})$ denotes photon survival
 probability only under the absorption of the Gilmore EBL,
 $P_{\rm GC}$=exp$(-\tau_{\rm GC})$ presents the absorption of
 the combined Gilmre and CIBER EBL, and $P_{\rm
 GCA/GA}$ is equal to $P_{\gamma\rightarrow\gamma}$ with/without CIBER EBL. X represents PL, LP or PLC. The model with ALP have two additional free parameter, i.e.,~$g_{a\gamma}$ and $m_{a}$, relative to the other two models.

 \emph{Fitting}---For any given observed spectrum (HE+VHE) and absorption model, e.g.~for PKS 1424+240(2009) and $P_{\rm GCA}$ with given $g_{a\gamma}$ and $m_{a}$, we need to try fitting the observed spectrum with the three theoretical spectra respectively, and the function $\phi_{\rm X}$, e.g.~$\phi_{\rm LP}$, will be chosen among the four functions when its theoretical spectrum, e.g.~$\phi_{\rm GCA, LP}$, achieves the best fit (according to average chi-square value per degree of freedom) \cite{HESS2013, Biteau2015}.

 We perform a fit to the observed spectra with the four
 theoretical spectra $\phi_{\rm G/GC/GCA/GA, X}$ respectively, minimizing the $\chi^{2}$ function
 which allows to quantify the goodness of fit and obtaining the best-fitting parameters including those of the intrinsic spectra and the ALP parameters in the GCA/GA model.
 In this process, the ALP parameters for every spectrum-fitting are in common and in the parameter region $(m_{a}, g_{12}) \in[0.1\,\rm neV, 2.5\,neV]\times[2.9, 40]$. It can be probed by the planed IAXO experiment and the part of $g_{12}<5$ is outside the excluded region constrained by gamma-ray observations (SN1987A and Fermi/LAT NGC1275).

\emph{Hypotheses test and significance level}---In order to compare the fitting results from four different models and find the best model, we built the following hypotheses:
\begin{enumerate}[(i)]
\item
\boldsymbol{$H_{0}$}:~the CIB is entirely explained by the Gilmore EBL model and fit with function $P_{\rm G}$.

\item
\boldsymbol{$H_{1}$}:~the excess CIBER EBL is a new EBL component and no-ALP fit with function $P_{\rm GC}$.
\item
\boldsymbol{$H_{2}$}:~the excess CIBER EBL is/isn't a new EBL component and ALP exist, fit with functions $P_{\rm GCA/GA}$.
\end{enumerate}
Beside testing $H_{0}$ vs $H_{1}$ and $H_{2}$, we will also test $H_{1}$ vs $H_{2}(\rm GCA)$ to study the effect of ALP/photon coupling. Firstly, we test the hypotheses by comparing the statistic of minimum $\chi^{2}/\rm dof$ among different models. If the statistic of alternative hypotheses is smaller, then we will construct F-test statistic to calculate the significance level of rejecting null hypotheses, that is:
\begin{equation}
\label{eq:ftest}
f:=\frac{(\chi^2_{H'}-\chi^2_{\rm
H''})/(m-k)}{\chi^2_{H''}/(n-m)}\sim F_{m-k,n-m}.
\end{equation}
where $n$ denote the sample size, $k$ and $m$ are the number
of freedom parameters for the null hypothesis $H'$ and alternative hypothesis $H''$ respectively. This statistic follows the $F$-distribution with $n-m$ degrees of freedom in the denominator and $m-k$ degrees of freedom for the summed chi-squares in the nominator \cite{Majumdar2018}.

 The procedure we perform for the conversion scenario a) with negligible oscillation in IGMF will be presented in the subsection D of Sec.~\ref{sec:results}.

\section{Tesults and discussion}
\label{sec:results}
\begin{figure*}[t]
\centering
\includegraphics[width=0.4\textwidth,height=0.3\textwidth]{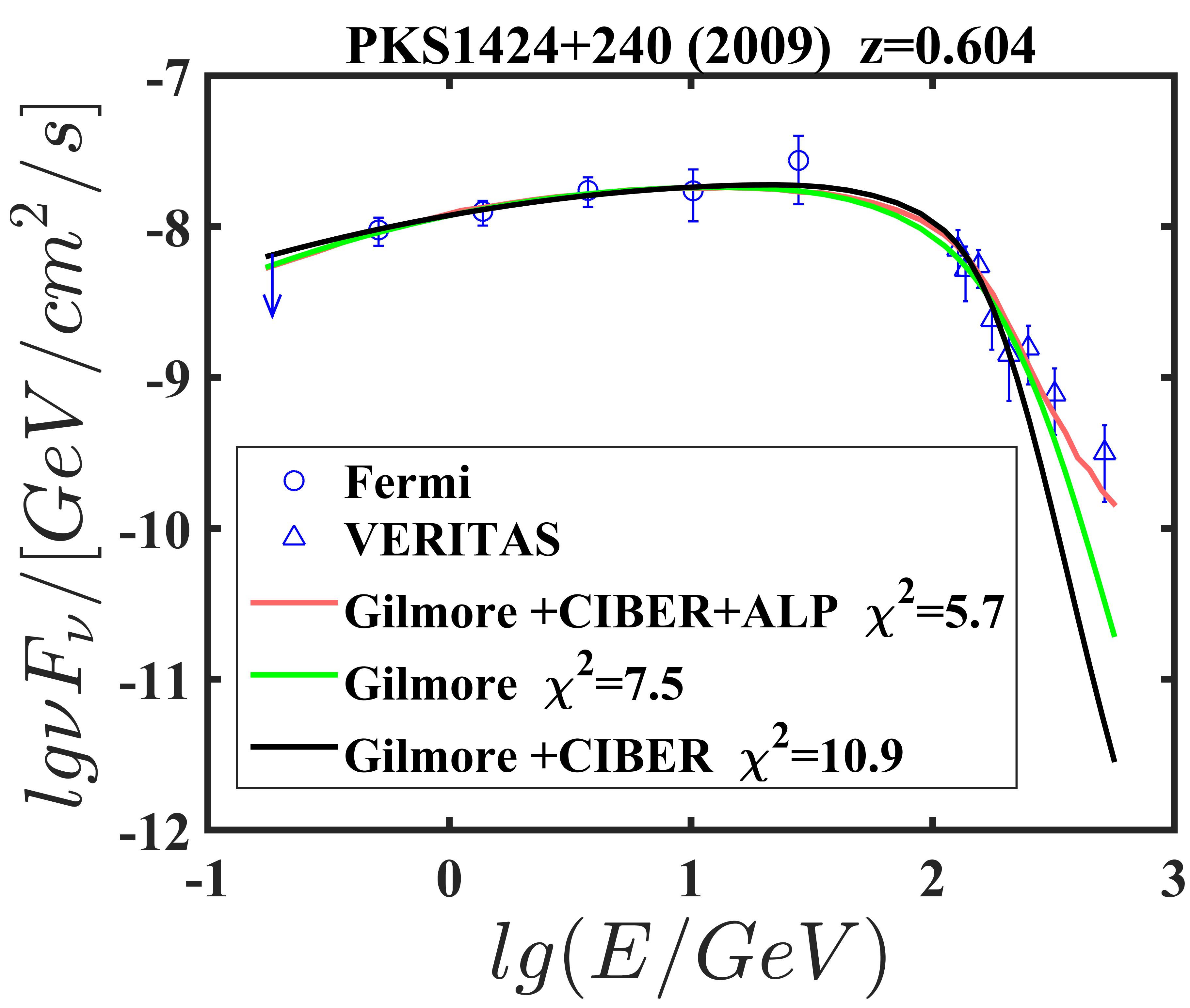}
\includegraphics[width=0.4\textwidth,height=0.3\textwidth]{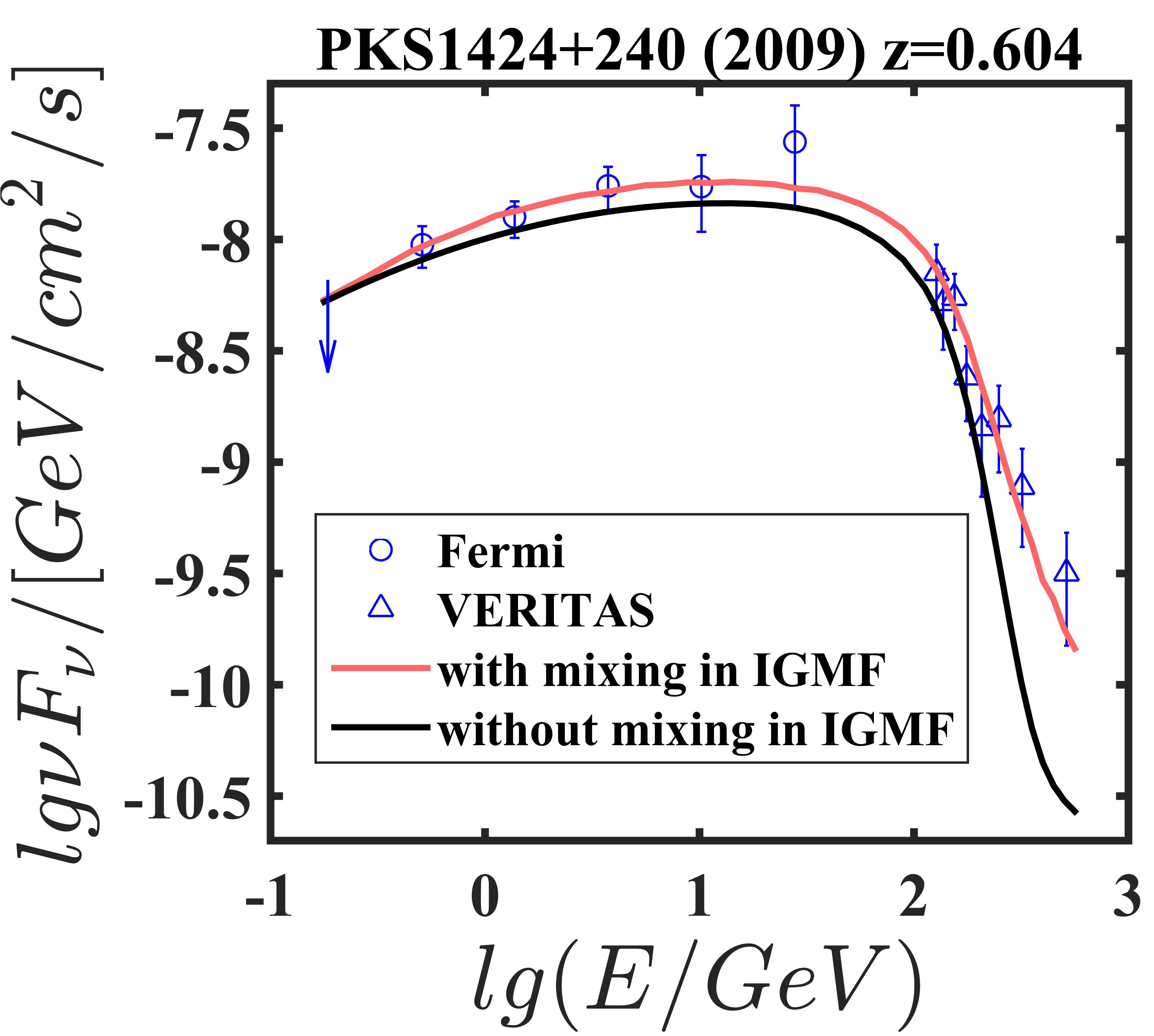}
\caption{Left panel: Fitting to the spectrum of PKS 1424+240
with the theoretical spectra ($\nu F_{\nu}=\nu^2\phi_{\rm G/GC/GCA, LP}$). The best
fitting ALP-parameters are $g_{12}=4.9$ and
$m_a$=0.1\,neV. Right panel: the effect of extragalactic conversion
to the best fitting spectrum in the left panel.}
\label{fig:4}
\end{figure*}

The results (for the conversion scenario including IGMF) of the spectral analysis and fitting
of the four models to the spectral points, are summarized
in the last column of Table~\ref{tb1}, Fig.~\ref{fig:4}, Table~\ref{table:chi}, Fig.~\ref{fig:5} in Appendix A.

The last column of Table~\ref{tb1} gives the chosen models for the intrinsic spectrum. The four absorption models almost have the same form of the best fitting intrinsic functions for each observed spectrum. This reduces the effect due to the different of intrinsic spectrum on the comparison of the absorption models.
\subsection{The best-fitting theoretical spectra}
Fig.~\ref{fig:4} (left) shows the SED of PKS 1424+240(2009),
overlaid with the three best-fitting spectra according to Eq.(\ref{eq:model}): GEBL (green line), GC (black line) and GCA(red line). Fig.~\ref{fig:4} (right) reveals that the ALP/photon conversion in extragalactic medium plays an important role in shaping the harness of theoretical spectrum at VHE. Obviously the GC model cannot fit the VHE spectrum well, and the GCA model achieve an acceptable fitting. This improvement to the goodness of fit ($\Delta \chi^{2}$) can be reflected quantitatively from the resulting $\chi^{2}$:~achieve with $\Delta\chi^{2}=4.78$ relative to that from GC model spectrum. Upon closer inspection of the SED, the improvement is a result of photon/ALP conversion in jet and extragalactic space during the beam propagation, where a part of HE photons damping due to the conversion but partial VHE photons avoid to be absorbed by the EBL soft photons and lead to a hard spectrum at VHE. A similar situation can be seen in Fig.~\ref{fig:5} except for the spectra 1ES 0347-121 , 1ES 0229(2009) and PKS 2155-304(2008). In these spectra the goodness of fit can not be improved for $m_{a}$=0.1 neV and $g_{12}$=4.9. Note that, this modification by the photon/ALP mixing is directly linked to the strength of the transversal magnetic field along the line of sight and the jet parameters.

For 1ES 1011+496 and 1ES 1218+304, the theoretical spectra $\phi_{\rm GCA, X}$ can not fit the observed spectra well even though the goodness of fit is significantly improved in comparison to the GC model. As a result, these two sources contribute about a third of the $\chi^2$ of the models including CIBER EBL, see Table~\ref{table:chi}. This is mainly because the effect of excessive absorption on the spectra at hundreds of GeV by the CIBER EBL photon can not offset effectively through the photon/ALP mixing. It is worth emphasizing that the data quality of these sources is relatively high.

 Compare to the GEBL model, the GCA model spectra achieve a significant better fit to only three spectra, i.e.~PKS 1424+240, 1ES 1101-232, and the former achieve a acceptable fit for each spectrum, see Fig.~\ref{fig:4} and Fig.~\ref{fig:5}.

\begin{table*}[t]
\centering
\caption{A comparison of the minimum chi-square values
resulted from four different models and they are obtained
for the four hypotheses:~$H_{0}$(Gilmore), $ H_1$(Gilmore+CIBER), $H_2$(Gilmore+CIBER+ALP for the best-fitting values of $m_{a}$=0.1\,neV and $g_{12}$=4.9) and $H_2$ (Gilmore+ALP for the best-fitting values of $m_{a}$=0.5\,neV and $g_{12}$=2.9). The significance of improvement in the goodness of fit due to the introduction of ALP is calculated with the excess variance technique \cite{Majumdar2018}.
\label{table:chi} }
\begin{ruledtabular}
\begin{tabular}{lrrrrccccc}
   \multirow{2}{*}{Source} & $\chi_{\rm G}^{2}(\rm dof)$ &
   $\chi_{\rm GC}^{2}(\rm dof)$ & $\chi_{\rm GCA}^{2}(\rm dof)$ &
    $\chi_{\rm GA}^{2}(\rm dof)$ & significance \\
   & $H_{0}$ & $H_{1}$ & $H_{2}(\rm GCA)$ & $ H_{2}(\rm GA)$ &
    $H_{2}(\rm GCA)/H_{1}$ \\
\hline
PKS 2155-304(2008)& 9.81(9) & 5.22(9) & 5.74(9) &
 10.50(9) & - \\[3pt]
PKS 2155-304(2013)& 34.96(23) & 34.96(23) & 32.13(23) &
 34.23(23) & - \\[3pt]
1ES 0229+200(2005) & 10.48(8) & 11.60(8) & 10.56(8) & 10.74(8) & - \\[3pt]
1ES 0229+200(2009) & 5.30(10) & 8.20(10) & 8.80(10) & 5.11(9)\footnotemark[1] & - \\[3pt]
1ES 1218+304 & 15.24(12) & 49.52(12) & 23.10(12) & 16.10(12) & - \\[3pt]
1ES 1101-232(2004) &8.64(16)&9.28(16)& 7.42(16) & 8.12(16) & - \\[3pt]
1ES 0347-121 & 2.94(7) & 3.78(7)& 6.35(7) & 4.00(7) & - \\[3pt]
1ES 1011+496 & 14.52(11) & 59.84(11) & 25.74(11) & 18.09(11) &  - \\[3pt]
1ES 0414+009 & 4.97(7) & 9.87(7) & 4.10(7) & 5.05(7) &  - \\[3pt]
OT081 & 8.55(9) & 10.62(9) & 9.66(9) & 7.63(9) &  - \\[3pt]
PKS 0477-439& 15.10(10)& 27.50(10) & 17.84(10) & 12.96(10) & - \\[3pt]
PKS 1424+240(2009) &7.50(10) & 10.90(10) & 5.68(10)  & 4.96(10) & - \\[3pt]
PKS 1424+240(2013)&14.88(12) & 22.92(12) & 11.70(12) & 13.20(12) & - \\[3pt]
\hline
 Combined & 152.89(144) & 264.21(144) & 168.82(142) & 151.29(141) & 7.60$\sigma$ \\
$\chi^{2}/(\rm dof)$ & 1.06 & 1.83 & 1.19 & 1.07 & - \\[3pt]
\end{tabular}
\end{ruledtabular}
\footnotetext[1]{The best-fitting intrinsic spectrum is LP in this model while other models choose PL, so that the DOF here is 9, not 10.}
\end{table*}

\subsection{the best model}
The combined data of the resulting minimum
 $\chi^2$ value is calculated and the hypotheses $H_{1}$ and $H_{2}(\rm GCA)$ is tested with the statistic $F$, see Table.~\ref{table:chi}.
The overall fit is improved from the combined minimum $\chi_{\rm GC}^{2}=264.21$ to $\chi_{\rm GCA}^{2}=168.82$.
The maximal statistical significance under the hypotheses is $7.6~\sigma$ corresponding to the best-fitting values of $m_{a}$=0.1\,neV and $g_{12}$=4.9, and the minimum one approaches $2~\sigma$ for the worst-fitting values $m_{a}$=2.5\,neV and $g_{12}$=10. Therefore, the $H_{1}$ hypothesis can be rejected at 95\% C.L.

The GA ($H_{2}$)and GCA ($H_{2}$) model can fit the observations well overall with $\chi_{\rm GA}^{2}/dof$=1.07 and $\chi_{\rm GCA}^{2}/\rm dof$=1.19 respectively, see Table.~\ref{table:chi}. But both of them are larger than that ($\chi_{\rm G}^{2}/\rm dof$=1.06) from the GEBL model ($H_{0}$). Therefore, we accept the null hypothesis $H_{0}$ for any values in the region $(m_{a}, g_{12}) \in[0.1\,\rm neV, 2.5\,neV]\times[2.9,40]$. This result demonstrates that the gamma-ray observations of distant blazars prefer the null hypothesis $H_{0}$, i.e.,~the Gilmore EBL model.

If we only considered the hard spectra PKS 1424+240
and 1ES 1101-232, almost the entire region $(m_{a}, g_{12})\in[0.1\,\rm neV,~2.5\,neV]\times[2.9,40]$ is favored at 95\% C.L. with respect to the $H_{0}$ hypothesis. However, as pointed out by Biteau et al.\ \cite{Biteau2015},
it is possible that the so called spectral hardness or even
upturn at the highest energy appeared in some special blazars is resulted from leaving out the systematical uncertainty in the analysis. If taking it into account in the calculation of $\chi^{2}$, the theoretical spectra with GEBL should fit the observations better.
\subsection{Discussion}
In this subsection, we will mainly discuss the ALP/photon conversion (a) introduced above, the minimal
CIBER EBL, reasonableness about testing the lower bound of the nominal CIBER EBL, other possible solutions to the tension among the observational data, systematic uncertainties and data quality.

\begin{table*}[t]
\centering
\caption{The goodness of fit for the conversion scenarios (a):~ALP/photon conversion occurs in the gamma-ray source and then further reconvert in the MW and $(m_{a},g_{11})\in[1\,\rm neV,100\,neV]\times[1,6.6]$. Base on the minimum $\chi^2$
resulted from four different models, we test
the four hypotheses:~$ H_{0}$(Gilmore), $ H_1$(Gilmore+CIBER), $H_2$(Gilmore+CIBER+ALP for the best-fitting values of $m_{a}$=1\,neV and $g_{11}$=1.5) and $H_2$ (Gilmore+ALP for the best-fitting values of $m_{a}$=1 neV and $g_{11}$=1.5). The minimum CIBER EBL is also tested. The significance of improvement in the goodness of fit due to the introduction of ALP is calculated with the excess variance technique.
\label{tb4} }
\begin{ruledtabular}
\begin{tabular}{lrrrrccccc}
   \multirow{2}{*}{Source} & $\chi_{\rm GC}^{2}(\rm dof)$ \footnotemark[1] &
   $\chi_{\rm GCA}^{2}(\rm dof)$ & $\chi_{\rm GA}^{2}(\rm dof)$ &
    $\chi_{\rm min}^{2}(\rm dof)$ & significance \\
   & $H_{1}$ & $H_{2}(\rm GCA)$ & $H_{2}(\rm GA)$ & minimum
CIBER &
    $H_{2}(\rm GCA)/H_{1}$ \\
\hline
PKS 2155-304(2008)& 4.86(9) & 4.62(9) & 13.16(9) &
 34.50(23) & - \\[3pt]
PKS 2155-304(2013)& 34.96(23) & 30.03(23) & 31.50(23) &
 4.77(9) & - \\[3pt]
1ES 0229+200(2005) & 10.96(8) & 9.60(8) & 10.50(8) & 11.52(8) & - \\[3pt]
1ES 0229+200(2009) & 7.00(10) & 7.68(10) & 6.72(9)\footnotemark[2] & 5.70(9) & - \\[3pt]
1ES 1218+304 & 42.24(12) & 39.90(12) & 17.90(12) & 22.68(12) & - \\[3pt]
1ES 1101-232(2004) &8.64(16)&8.16(16)& 7.42(16) & 7.52(16) & - \\[3pt]
1ES 0347-121 & 3.43(7) & 3.40(7)& 6.35(7) & 4.06(7) & - \\[3pt]
1ES 1011+496 & 50.82(11) & 63.00(11) & 15.21(11) & 29.37(11) &  - \\[3pt]
1ES 0414+009 & 9.38(7) & 6.80(7) & 3.40(7) & 6.79(7) &  - \\[3pt]
OT081 & 10.44(9) & 9.66(9) & 7.98(9) & 10.53(9) &  - \\[3pt]
PKS 0477-439& 26.80(10)& 24.00(10) & 20.00(10) & 23.30(10) & - \\[3pt]
PKS 1424+240(2009) &10.07(10) & 8.16(10) & 4.24(10)  & 9.70(10) & - \\[3pt]
PKS 1424+240(2013)&21.84(12) & 12.8(12) & 11.50(12) & 18(12) & - \\[3pt]
\hline
 Combined & 241.83(144) & 224.23(142) & 151.63(141) & 188.44(144) & 2.60$\sigma$ \\
$\chi^{2}/(\rm dof)$ & 1.68 & 1.58 & 1.08 & 1.31 & - \\[3pt]
\end{tabular}
\end{ruledtabular}
\footnotetext[1]{Here, the lower limit of two $\sigma$ statistical uncertainty and a systematic error of the CIBER data are taken as done with Ref.~\cite{Kori2017}, so the resultant $\chi^{2}$ has small difference with the case a $\sigma$ statistical uncertainty is adopted in Table~\ref{table:chi}.}
\footnotetext[2]{The best-fitting intrinsic spectrum is LP in this model while other models choose PL, so that the dof here is 9, not 10.}
\end{table*}

  In the conversion scenario (a), the ALP/photon conversion in IGMF is negligible, whereas the conversion in ICMF would be taken into account so that more free parameters will be introduced. Nevertheless, for simplicity, we do not consider the structure of magnetic fields in the source, i.e., the magnetic fields are assumed to be homogeneous in the jet and in the cluster, thus the relevant conversion probability that the emitted gamma-ray photons are converted to ALPs can be calculated simply by Eq.~\ref{probability}. The size of conversion region of source is assumed to be 10\,kpc and the magnetic field $B_{\rm T}=10\,\mu \rm G$, which can satisfy the Hillas criterion for accelerating the ultrahigh-energy cosmic rays \cite{Kori2017, Hooper2007}. For an efficient conversion under this assumption, we consider the ALP-parameters region $(m_{a},g_{11})\in[1\,\rm neV,100\,neV]\times[1,6.6]$. The upper limit of $g_{a\gamma}$ are given by the constraints of the CAST experiment \cite{CAST2017}. The lower limit of $m_{a}$ is result from Eq.~\ref{Ecrit} in the condition of an inefficient conversion in IGMF for $B_{\rm T}\thicksim1\,\rm nG$ and the constrained by $\gamma$-ray \cite{Fermi2016} observations. Since the phase of the sine function in Eq.~\ref{probability} is larger than 1 rad for the considered $g_{a \gamma}$, $B_{\rm T}$ and r, i.e.,~$1.5\frac{g_{a \gamma}}{10^{-11}\,{\rm GeV}^{-1}}\frac{B_{\rm T}}{10\,\mu \rm G}\frac{r}{\rm 10\,kpc}\sqrt{1+(\frac{E_{\rm crit}}{E_{\gamma}})}>1$, we take the average value for the square of the sine function to smear out the rapid-oscillatory features of the probability function \cite{Kori2017, Hooper2007}.

 As done above, we perform a combined global fit to the spectra with the GCA and GA model without taking into account the conversion in IGMF. The results are shown in Table~\ref{tb4}. The goodness of fit between the models including CIBER EBL is improved only with a maximal significance of $2.6~\sigma$ due to the introduction of ALP. The minimum $\chi_{\rm GA}^2/\rm dof$ (1.08) and $\chi_{\rm GCA}^2/\rm dof$ (1.58)
 are larger than those (1.07 and 1.19) of the models taking into account the conversion in IGMF respectively. Therefore the observations for our sample are in favor of the conversion scenario with IGMF between the two mechanisms for the GC model. The goodness of fit with ALP is also inferior to that with Gilmore EBL model. So we still accept the null hypothesis $H_{0}$. Part of the reason for this difference is that the rapid-oscillatory features of the probability function for the conversion in JMF are smeared out.

  Compared to Ref.~\cite{Kori2017}, the $\chi^{2}$ values of fitting to the spectrum of 1ES 1101-232 with ALP-relevant model are comparable ($\chi^{2}/\rm dof\thicksim$ 0.6, for $m_{a}\thicksim1\,\rm neV$ and $g_{11}\thicksim3.5$), even though we consider the structure of GMF.

  The tension
between the minimal CIBER EBL and the gamma-ray spectra is worth to be discussed. We take average of the data and test it. The minimum $\chi_{\rm min}^2/\rm dof$ is 1.31, see Table~\ref{tb4}, and is larger than those of the GEBL and models with ALP. It is mainly in tension with the observations of 1ES 1218+304, 1ES 1011+496 and PKS 0447-439. The minimal CIBER EBL has been recently studied by Ref.~\cite{Korochkin2019}.

Reasons about testing the lower bound of the nominal CIBER EBL. The best-fit value of $g_{a\gamma}=4.9$ locates at the edge of the excluded region by observations. If we take the average value, then the higher $g_{a\gamma}$ value would be required to offset the increased EBL absorption, and the best-fit $g_{a\gamma}$ would fall into the excluded region. On the other hand, the absorption curve around 1\,TeV would become more concave. As a result, the challenge for the ALP/photon coupling mechanism to explain the observation or mimic the absorption function of GEBL model successfully would increase greatly. The relation $\chi_{\rm GCA}^2>\chi_{\rm GA}^2$ hints higher CIBER EBL is not preferred by the coupling mechanism. Therefore testing the lower bound could largely represent the whole nominal CIBER EBL to draw a consistent conclusion.

 An alternative and possible explanation for the spectral
 hardening (e.g.~PKS 1424+240 \cite{Zheng2013}, 1ES 1101-232
 \cite{Essey2013, Cerruti2017}) is the ``hadronic cascade
 model", in which an additional $\gamma$-ray emission
 component due to the interactions of EBL or CMB photon with
 proton Cosmic Rays originating from the blazar along the
 line of sight lead to the spectral hardening toward the
 highest energy \cite{Essey2010, Dzhatdoev2017}. However, it
 is only appropriate for some special sources with $z>0.15$
 and long time variability \cite{Prosekin2012, Galanti2015}, thus the
 combination of this mechanism and CIBER EBL cannot
 explain the $\gamma$-ray observations of all the blazars.
 Another model beyond traditional physics is lorentz
 invariance violation (LIV). But it may lead to a
 significant reduction of the $\gamma$ - $\gamma$ opacity
 for photons with energies $E$ $>$ 10 TeV and can only explain the observed spectra of some special blazars
 \cite{Abdalla2018, Tavecchio2016, Jacob2008}.

  We discuss the sources of systematic uncertainties of the study now. The photon survival probability
  $P_{\gamma\rightarrow\gamma}^{\rm JMF}$ are sensitive to the
  jet parameters $B_{\rm VHE}$ and $R_{\rm VHE}$, thus the
  uncertainties of these values have a significant impact on
  the total systematic uncertainties of the relevant result, mainly the best-fit ALP parameters. Therefore we do not aim at constraining the ALP parameters. The values of these jet parameters can be constrained in principle with multi-frequency observations and theoretical models.

  The uncertainty resulted from the choice of intrinsic
  spectral models and the uncertainty of the observed data
  from the telescope, e.g.~energy calibration, have an
  important contribution to the error of the result. The choice of the observed spectra included in the sample is biased by the criteria (e.g.~the highest intrinsic energy $>800$\,GeV) thus selecting preferentially objects with hard spectra, e.g.~the three hard spectra PKS 1424+240 and 1ES 1101-232. Some of them are usually hard to explain by the traditional absorption model e.g.\ Refs.~\cite{Abdalla2018, Aharonian2007, Essey2012} and seem to prefer the model with ALP. As a result, the sample is biased towards hypothesis $H_{2}$ (with ALP) compared to $H_{0}$. However this bias should not change our conclusion, as it prefers the hypothesis $H_{0}$ instead.

  It is not enough for 98 random realizations to reduce the uncertainty effectively in the stochastic simulation. But the curves of theoretical spectra at VHE almost displayed a smooth shape, which illustrates the mean of 98 realizations have already closed to convergence.

  The data quality is important for our study. Few of the $\chi^2/dof$ for the individual
spectrum is less than 0.5, e.g.~$\chi_{\rm G}^2/\rm dof=0.42$ for 1ES 0347-121, which usually means the model have over-constrained for this spectrum. The cause for this problem lies in fewer data points and larger data uncertainty overall. To eliminate this adverse effect, we analyze the 13 spectra on the whole leading to the combined $\rm dof>140$ and a robust result $2>\chi^2/dof>1$. Moreover, though our result suggests that we have no need to introduce the ALP and support the recent EBL model, it is based on our sample that partly reflects the ability of current instrumental measurement. The upcoming telescopes, e.g.~CTA and LHAASO, would provide more
high quality data including that from more distant sources. Our results would be further tested in the future with more precise data and information on intrinsic physic of the source such as the intrinsic spectrum and the magnetic field structures of jet.

\section{summary and conclusion}
\label{sec:conclusion}
ALP is expected to potentially resolve the tension between the VHE $\gamma$-ray observations of distant blazars and the CIBER EBL \cite{Kori2017}, see Fig.~\ref{fig:1}. We aim to probe whether the excess radiation of CIBER is a new EBL component, in the background that the recent EBL models are repeatedly tested in many literatures and are generally agreed with the $\gamma$-ray observations, see e.g.,~Refs.~\cite{Fermi2012,
 Gong2013, HESS2013, VERITAS2015, Biteau2015, MAGIC2016,
 Armstrong2017, Desai2019}. Hence, we build four absorption models, i.e.,~GCA (Gilmore EBL+CIBER+ALP), GA (Gilmore EBL+ALP), GC (Gilmore EBL+CIBER) and GEBL (Gilmore EBL).

 In the model including ALP, the ALP/photon beams cross four magnetic-field regions on their propagation path up to Earth: jet, intra-cluster, extragalactic space and MW galaxy, see Fig.~\ref{fig:2}. For an efficient conversion at VHE in IGMF and due to the observation constraints on ALP parameter space \cite{Montanino2017}, we consider the ALP parameter range $(m_{a}, g_{12}) \in[0.1\,\rm neV,
 2.5\,neV]\times[2.9,40]$ for which the conversion in intra-cluster could be neglected, see Fig.~\ref{fig:3}. The conversion, in IGMF obtained from large-scale cosmological simulations and in large scale coherent JMF, is important to shape the hardness of spectrum at VHE, see Fig.~\ref{fig:2} and Fig.~\ref{fig:4}.

  A combined global fit is performed to the observed spectra  with the absorption models together with the chosen intrinsic spectra, minimizing the $\chi^2$. Three types of the best-fitting spectra are show in Fig.~\ref{fig:4} and Fig.~\ref{fig:5}. The hypotheses involving the four absorption models are tested based on their combined minimum $\chi^2$ of the 13 spectra. The goodness of fit for the GC model can be improved with a significance of $7.6~\sigma$ if $m_a=0.1\,\rm neV$ and $g_{12}=4.9$, see Table~\ref{table:chi}. However, the goodness of fit either given by the GCA model or GA model is still inferior to that with GEBL, i.e.~$\chi_{\rm G}^2/\rm dof<\chi_{\rm GA}^2/\rm dof<\chi_{\rm GCA}^2/\rm dof$.

We also discuss the complementary conversion scenario: the ALP/photon conversion occurs in the gamma-ray source where the magnetic fields in the blazar jet and intra-cluster are treated as homogeneous and in one zone, and then back-conversion occurs in the Milky Way. We find the maximum improvement to the goodness of fit after introducing the ALP/photon mixing mechanism is significant with 2.6 $\sigma$, which is inferior to that in the scenario involving the IGMF, see Table~\ref{tb4}. The goodness of fit from the GEBL model is also the best among the four models.

Though four of the sample are traditional hard spectra so that the sample selection is biased toward the model with ALP, the result more prefer the GEBL model overall instead. Other possible mechanisms for explaining the spectral harding such as LIV and hadronic cascade of cosmic rays are unlikely generally compatible with the TeV $\gamma$-ray observations of extragalactic sources under the absorption of CIBER EBL.

 We conclude that the ALP/photon mixing mechanism can effectively alleviate the tension between the $\gamma$-rays observations and the CIBER data. In particular, the spectra of PKS 1424+240 and 1ES 1101-232 are best explained by this mechanism if the systematic uncertainties on the measured flux are neglected. However, the Gilmore EBL attenuation can, on the whole, explain the $\gamma$-ray observations of our samples, and it is more consistent with the observations than the model including ALP under our magnetic scenario. Hence, in a statistical sense, assuming the existence of the ALP to alleviate the tension is not required, and the CIBER excess over the EBL models is less likely to be a new EBL component.

 The upcoming CTA and LHAASO would provide more
 precise data on VHE $\gamma$-rays from distant blazar. On the other hand, detecting the NIR
 EBL by an instrument in deep space where zodiacal light
 foreground is absent or with precise multi-band fluctuation
 measurements would be possible in the future
 \cite{CIBER2017}. These two improvements will help
 to clarify the uncertainties related to the origin of CIBER EBL and test our conclusion.

\begin{acknowledgements}
We thank anonymous referees for useful comments and suggestions. The author is indebted to F. Vazza for supplying the data
sets of simulated extragalactic magnetic fields. We thank Y. F. Liang for comments on an earlier version of the draft. W.P.L. acknowledges support from the National Key Program for Science and Technology Research and Development (2017YFB0203300), the National Key Basic Research Program of China (No. 2015CB857001), and the NSFC Grant 11473053. P.H.T is supported by the National Natural Science Foundation of China (NSFC) Grants 11633007, 11661161010, and U1731136. W.S.Z. is supported by the NSFC Grant 11673077 and 11733010.
\end{acknowledgements}

\clearpage

\widetext
\appendix
\section{fitting Results for the other 12 spectra}
\label{appendix}
\begin{figure*}[b]
\centering
\includegraphics[width=0.3\textwidth]{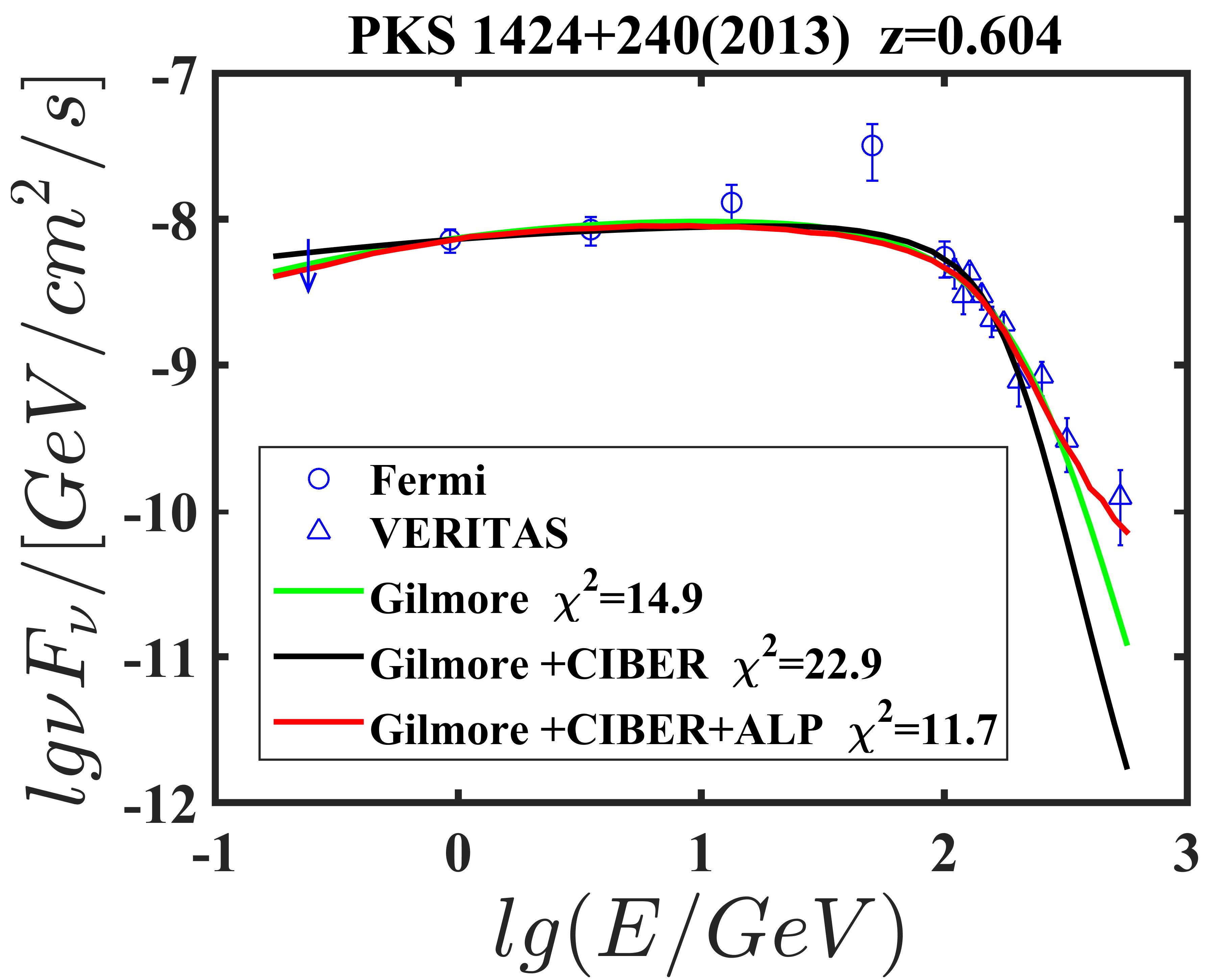}
\includegraphics[width=0.3\textwidth]{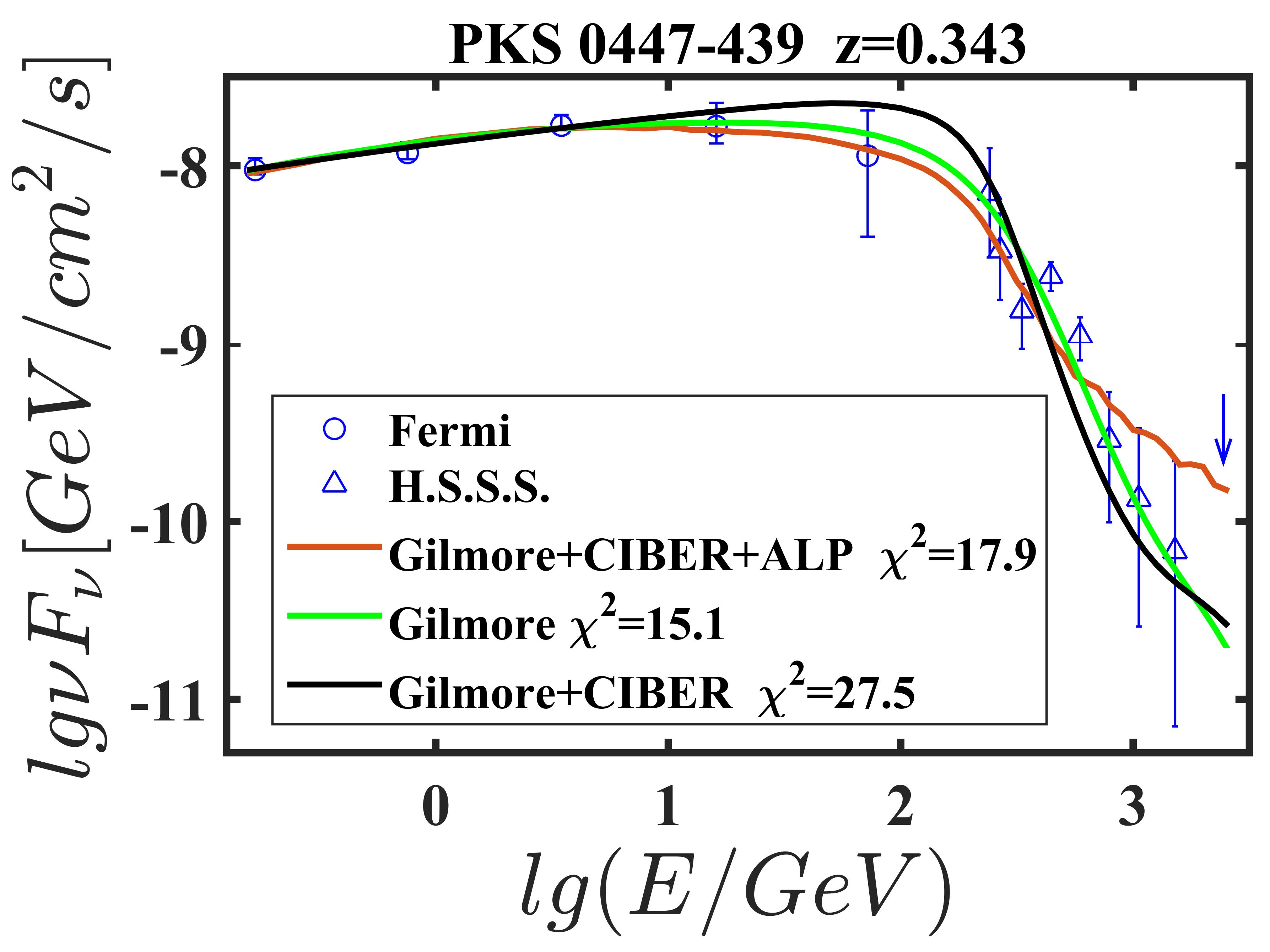}
\includegraphics[width=0.3\textwidth]{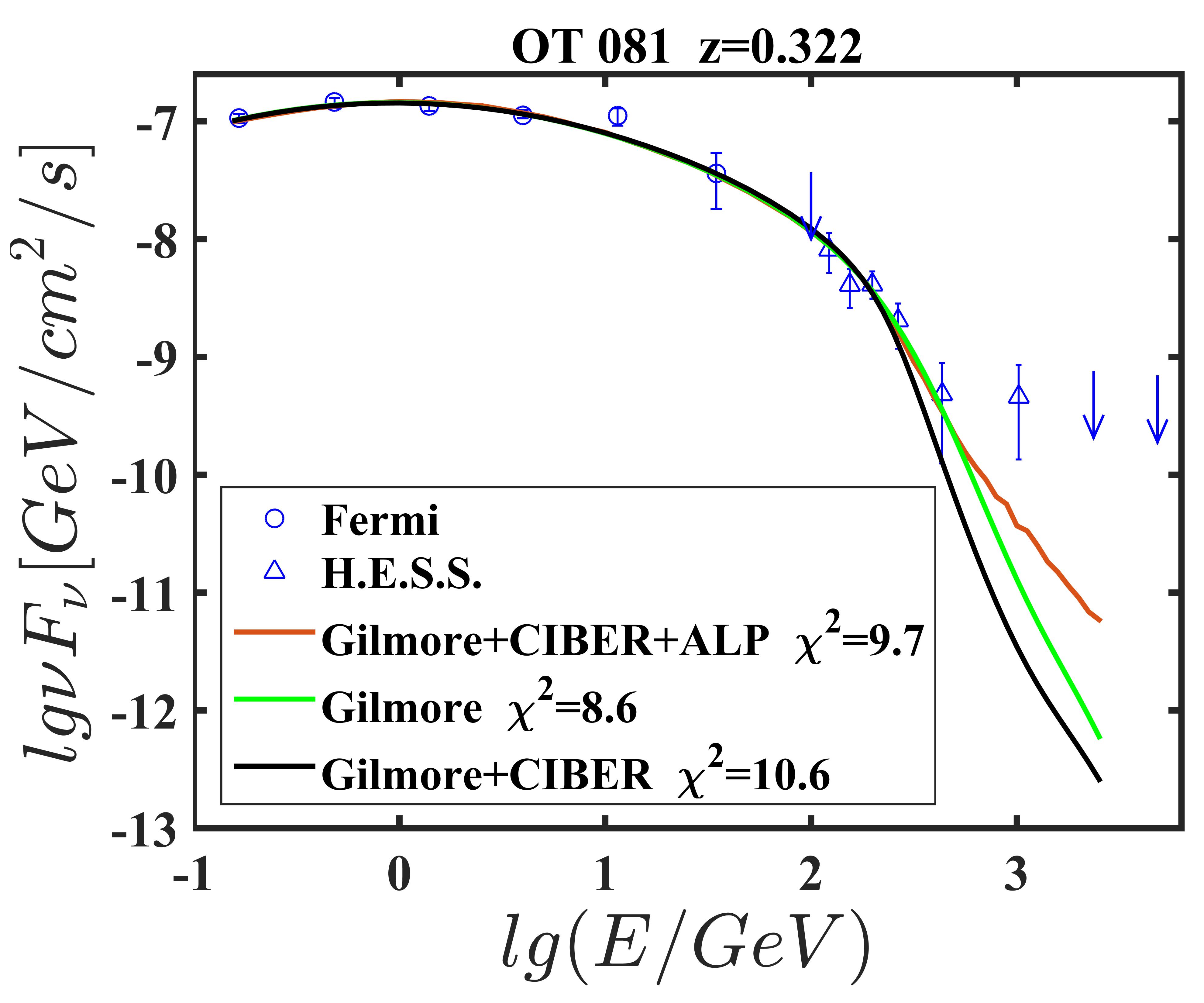}

\includegraphics[width=0.3\textwidth]{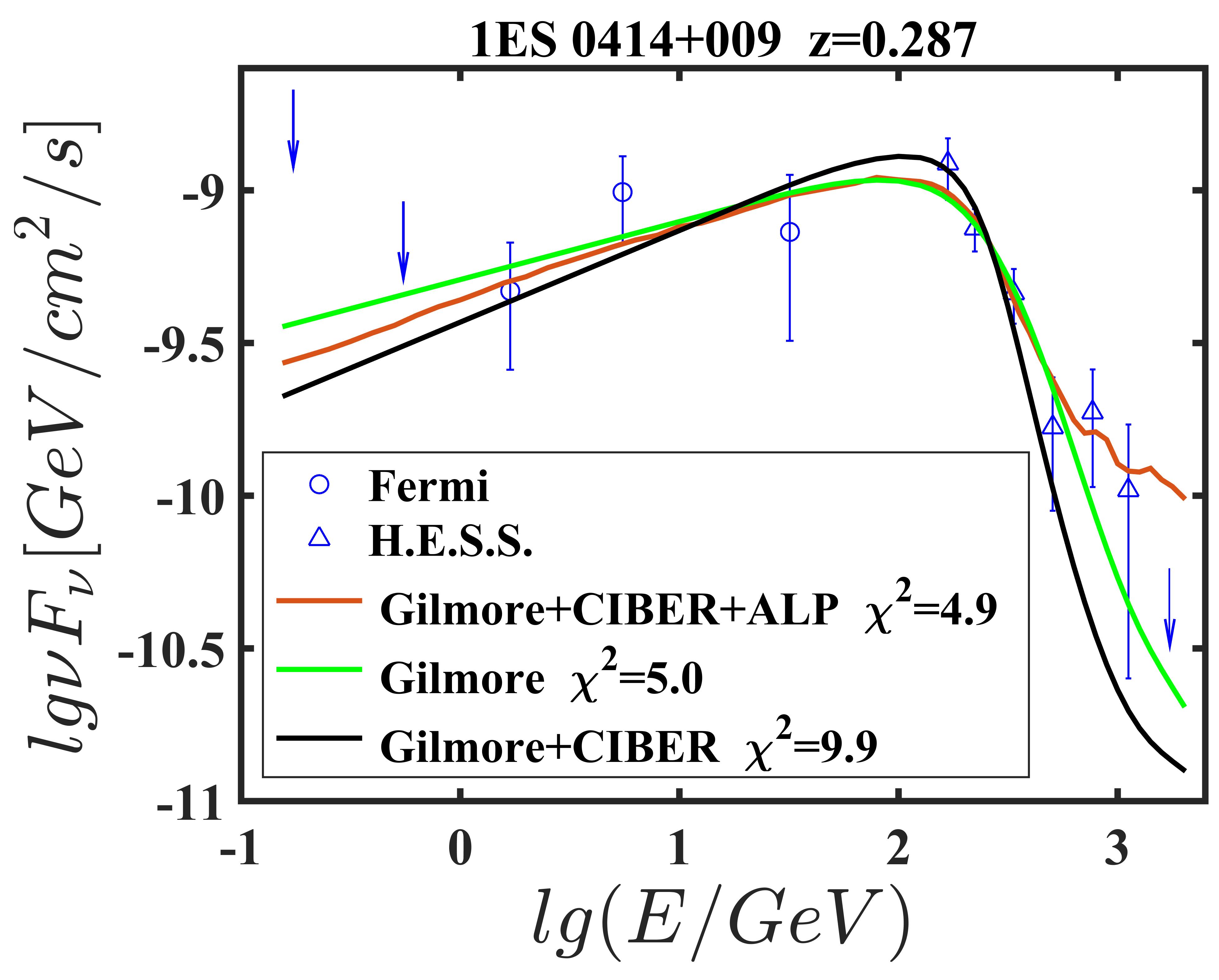}
\includegraphics[width=0.3\textwidth]{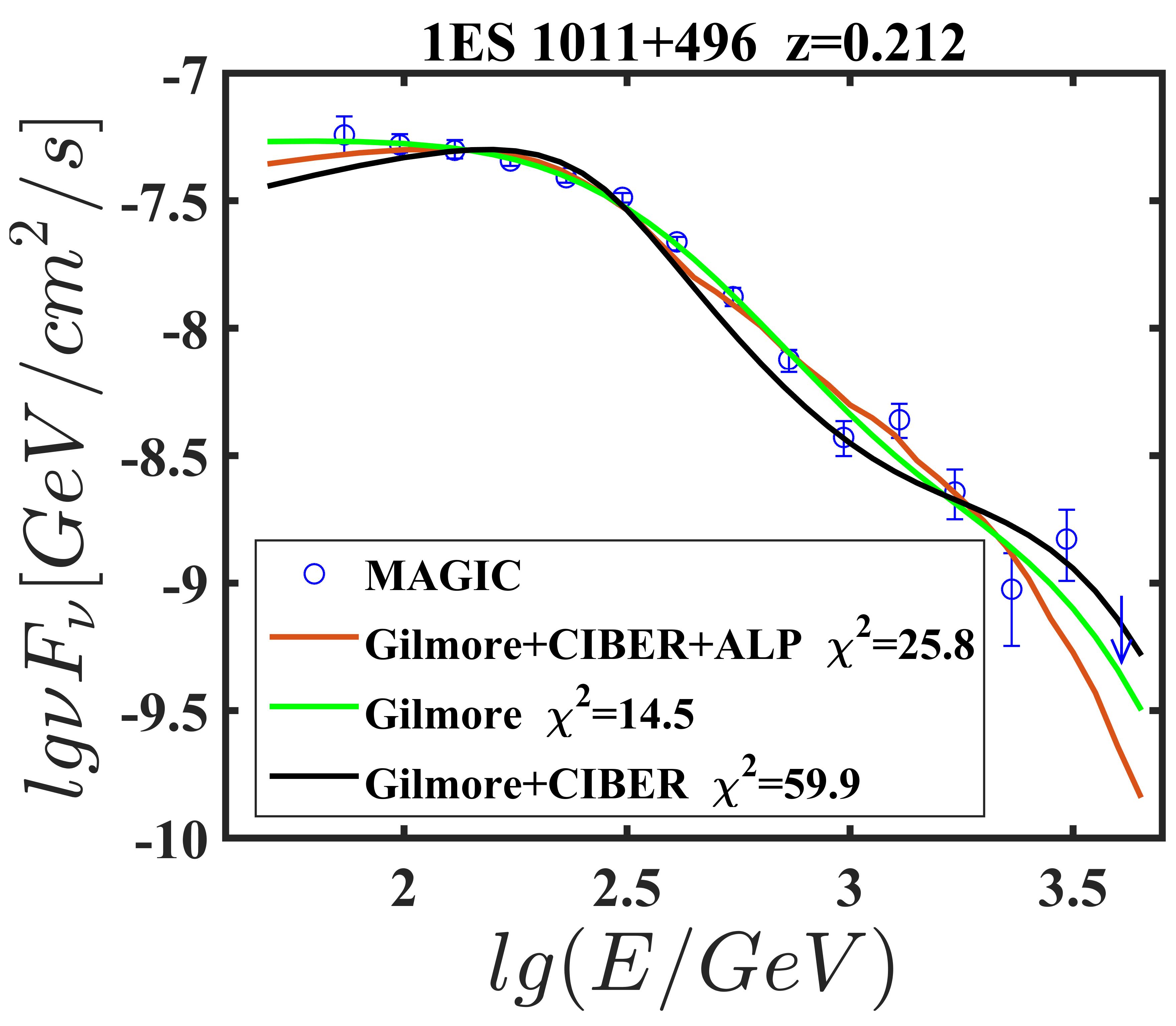}
\includegraphics[width=0.3\textwidth]{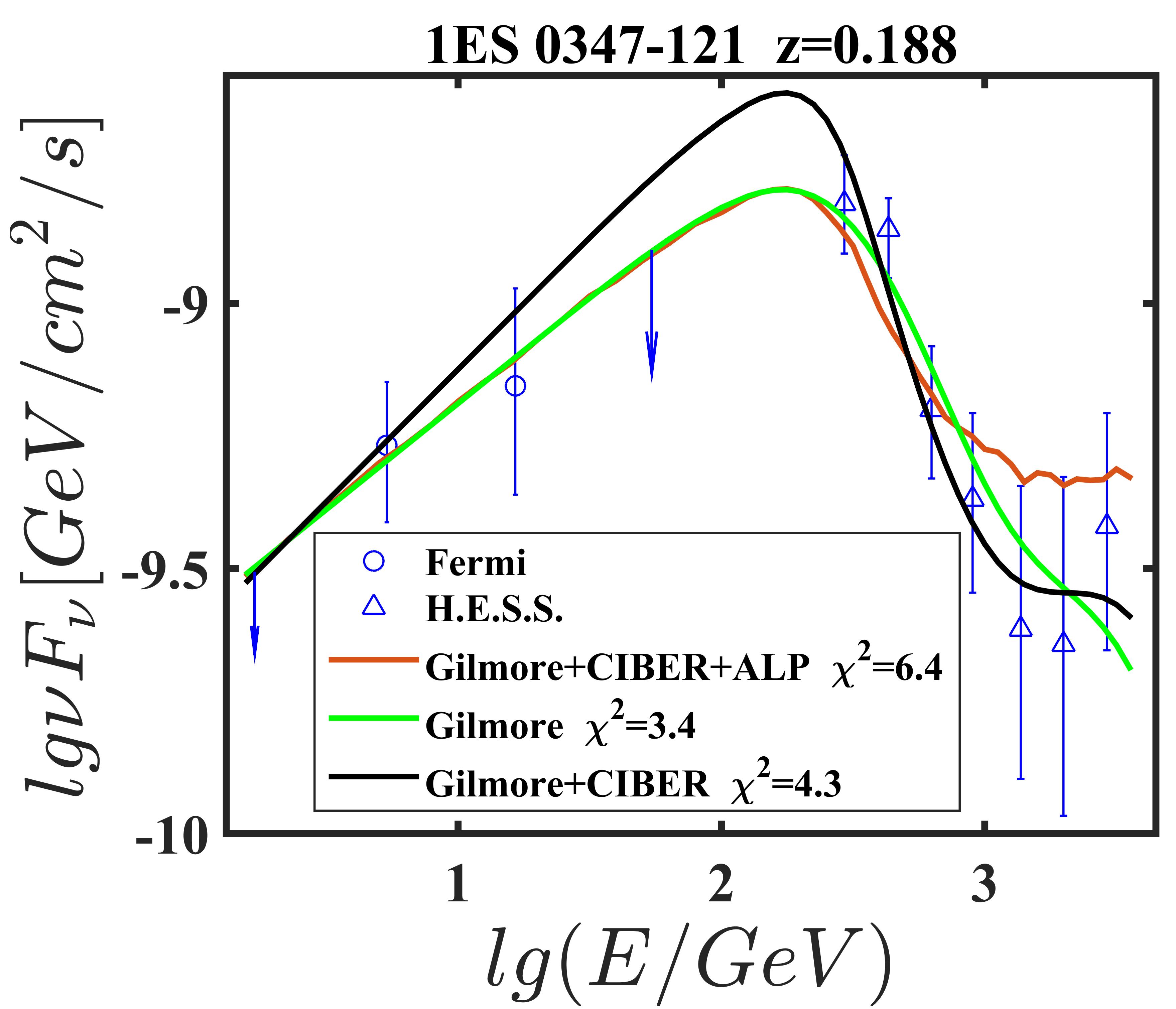}

\includegraphics[width=0.3\textwidth]{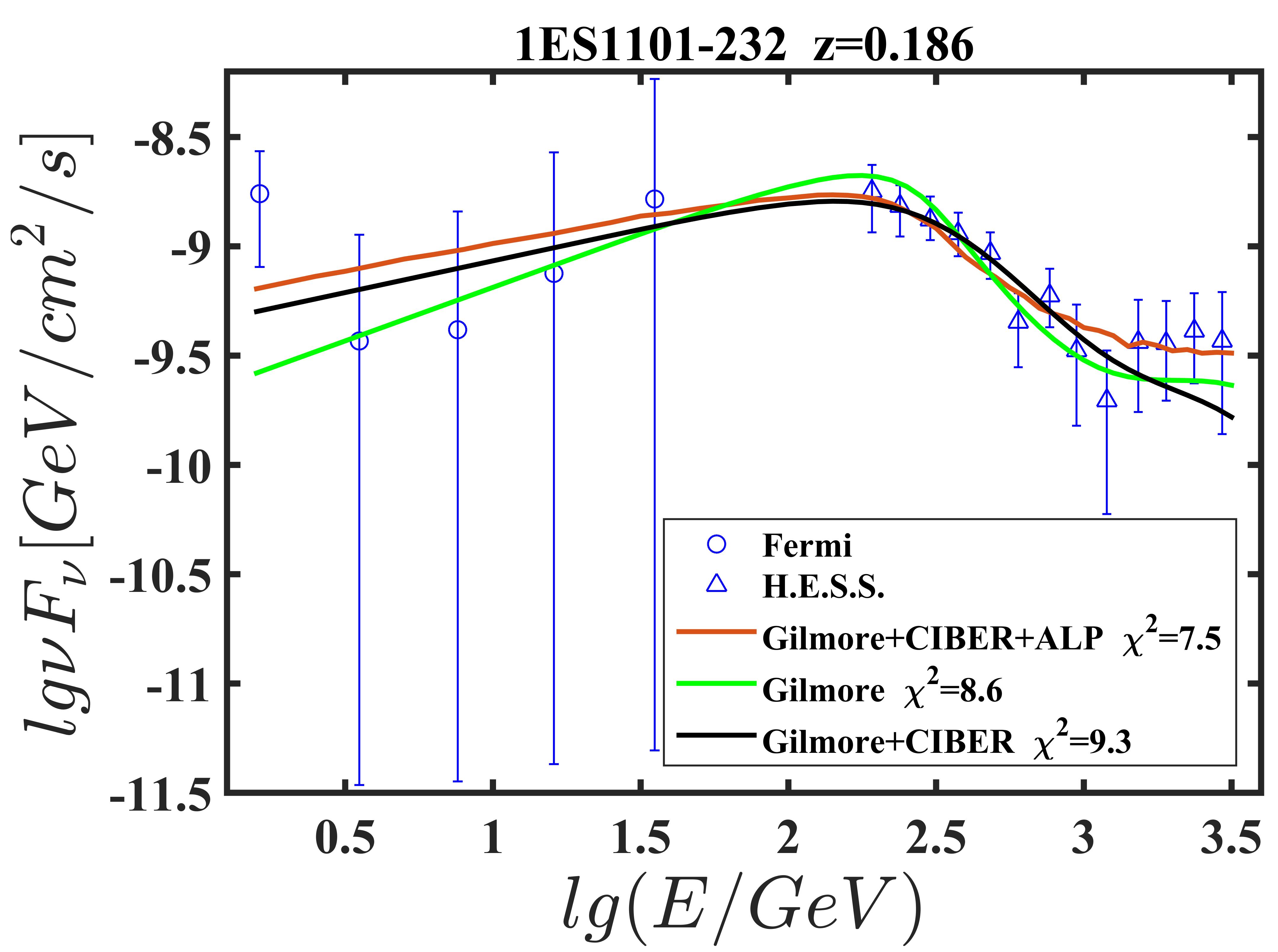}
\includegraphics[width=0.3\textwidth]{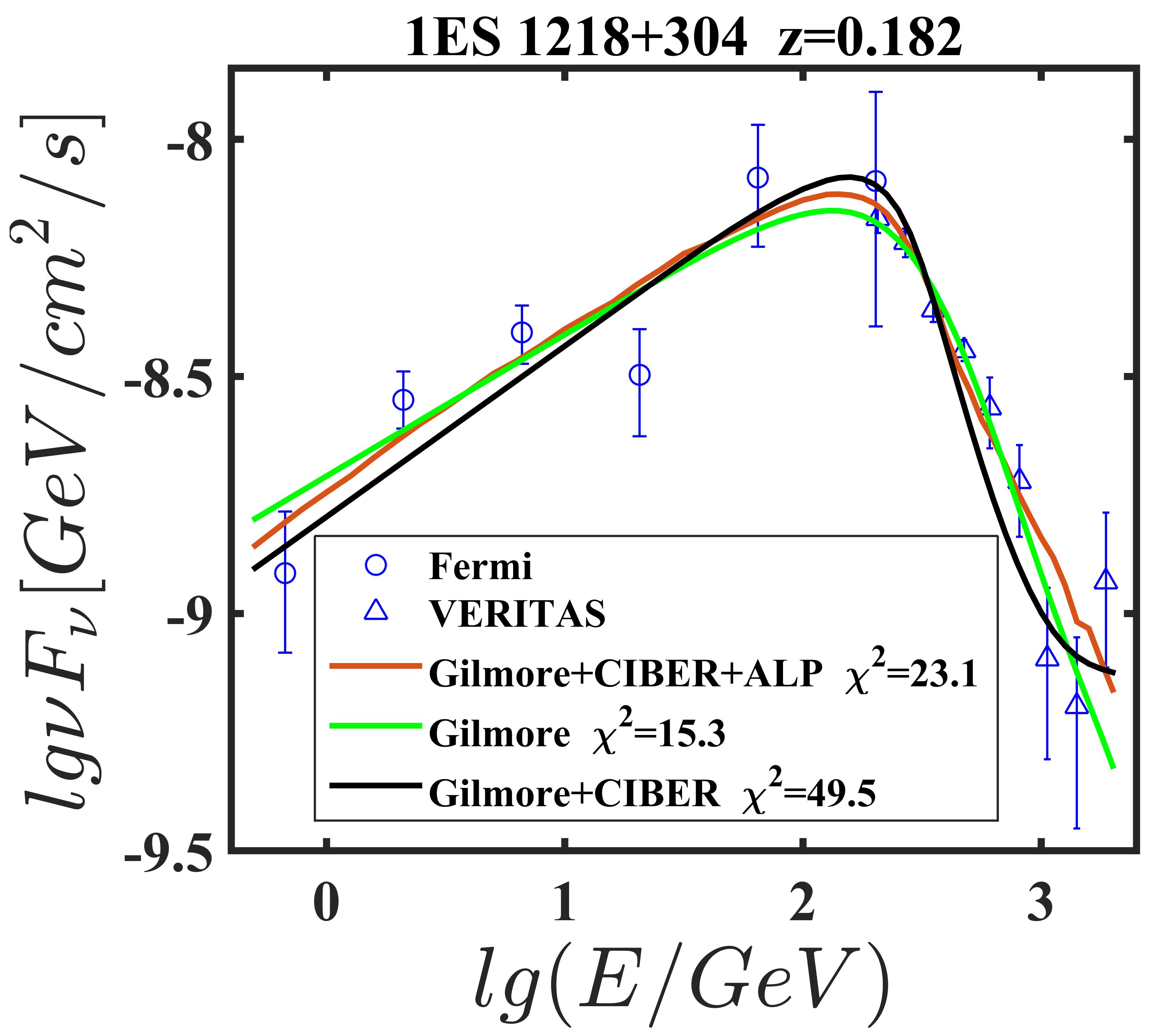}
\includegraphics[width=0.3\textwidth]{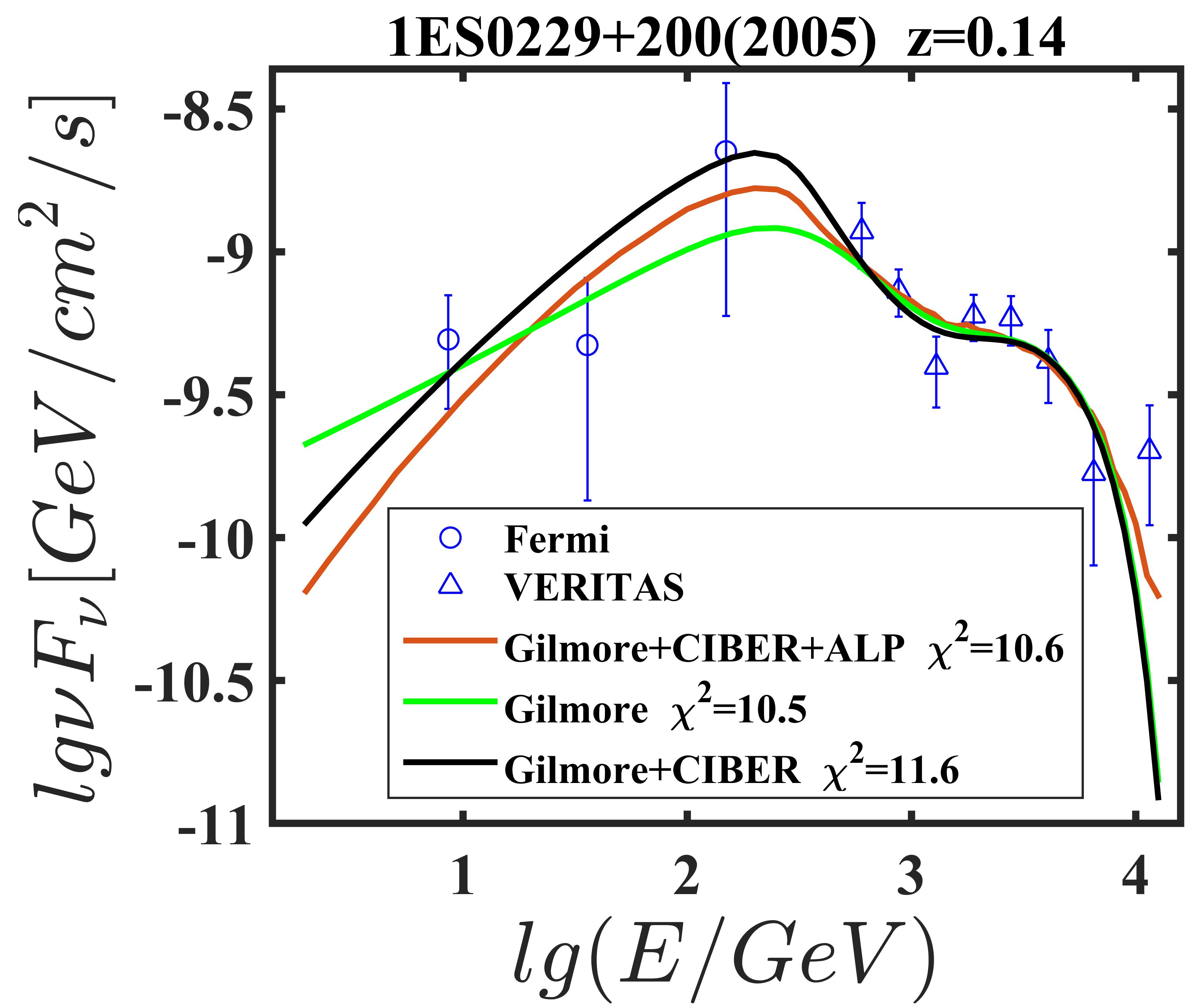}

\includegraphics[width=0.3\textwidth]{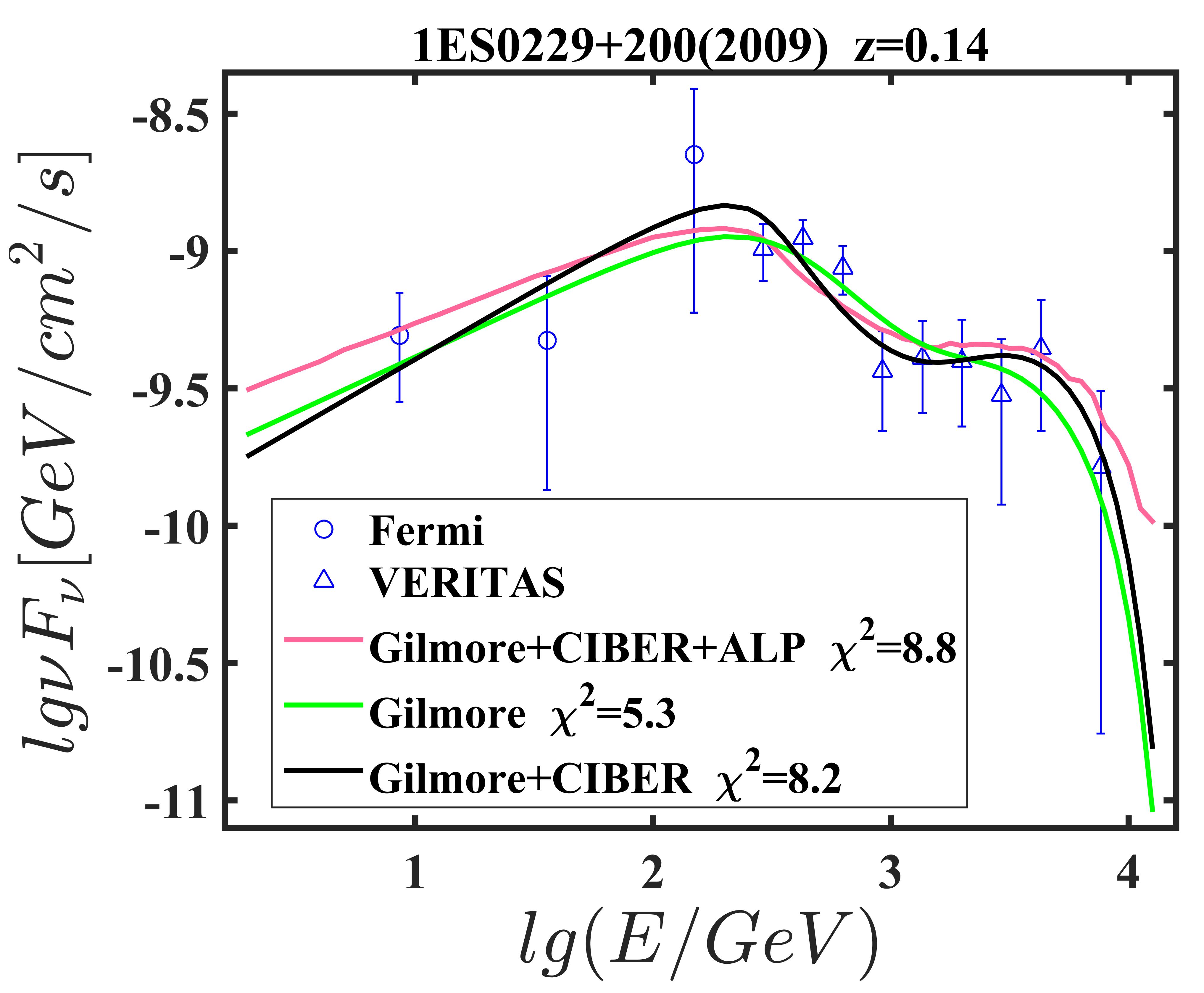}
\includegraphics[width=0.3\textwidth]{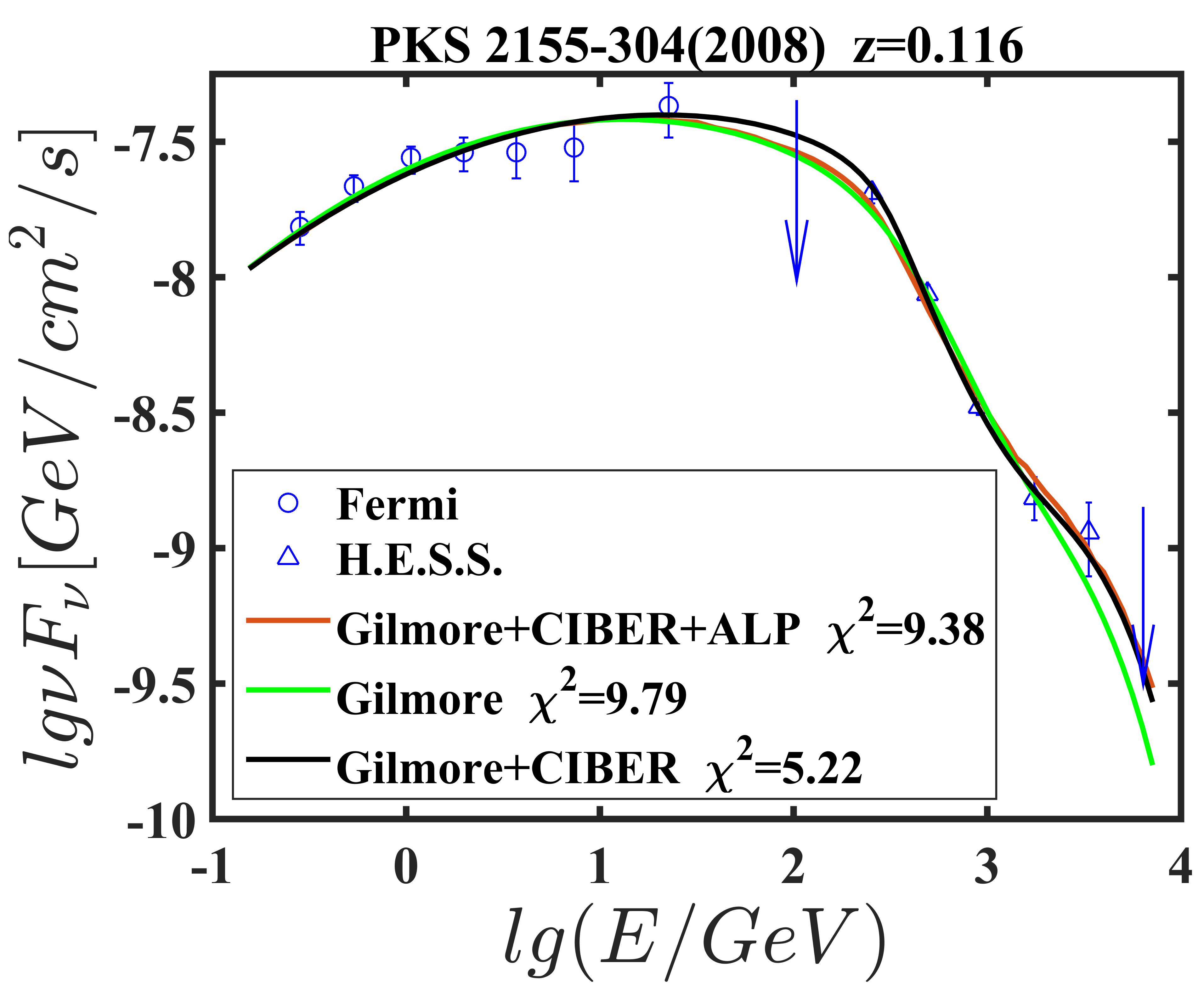}
\includegraphics[width=0.3\textwidth]{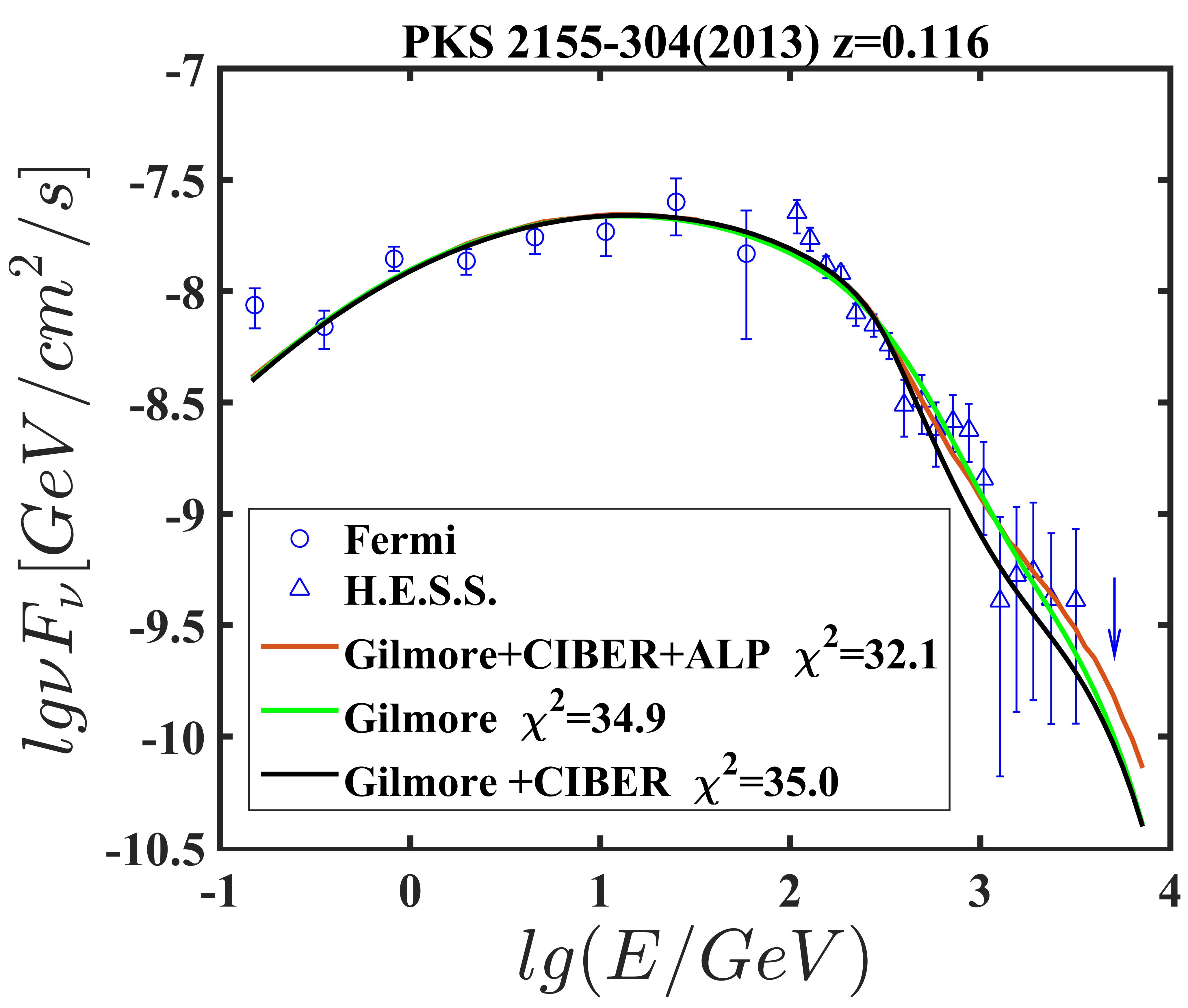}
\caption{Fitting to the observed spectra with theoretical
spectra. The best fitting ALP-parameter for each spectra
are $m_a$=0.1~neV and $g_{12}$=4.9.}
\label{fig:5}
\end{figure*}

\clearpage

\end{document}